\documentclass{article}
\usepackage{comment,amsmath,amsthm,amssymb,array,latexsym,cite,epsfig}
\usepackage{finalT}

\numberwithin{equation}{section}

\input pn.def

\begin{document}


\title{Post-Newtonian expansions for perfect fluids
}
\author{Todd A. Oliynyk\thanks{todd.oliynyk@sci.monash.edu.au} \\
School of Mathematical Sciences\\
Monash University, VIC, 3800\\
Australia}
\date{}
\maketitle

\begin{abstract}
\noindent We prove the existence of a large class of dynamical solutions
to the Einstein-Euler equations that have a first post-Newtonian expansion.
The results here are based on the elliptic-hyperbolic
formulation of the Einstein-Euler equations used in \cite{Oli06}, which
contains a singular parameter $\ep = v_T/c$, where $v_T$ is a characteristic
velocity associated with the fluid and $c$ is the speed of light. As in \cite{Oli06},
energy estimates on weighted Sobolev spaces are used to analyze the behavior
of solutions to the Einstein-Euler equations in the limit $\ep\searrow 0$, and
to demonstrate the validity of the first post-Newtonian expansion as an approximation.
\end{abstract}

\sect{intro}{Introduction}

The Einstein-Euler equations, which govern a gravitating
perfect fluid, are given by
\eqn{EEeqnA}{
G^{ij} = \frac{8\pi G}{c^4} T^{ij}\,
\AND \nabla_{i} T^{ij} = 0,} where \eqn{EEdefsA}{ T^{ij} = (\rho +
c^{-2} p)v^i v^j + p g^{ij},  } with $\rho$ the fluid density,
$p$ the fluid pressure, $v$ the fluid four-velocity normalized
by $v^iv_i = -c^2$, $c$ the speed of light, and $G$ the
Newtonian gravitational constant. Defining
\eqn{epdef}{\ep = \frac{v_T}{c}} where $v_T$ is a typical speed associated with
the fluid,
the Einstein-Euler equations,
upon suitable rescaling \cite{Oli06}, can be written
in the form \leqn{EEeqn}{ G^{ij} = 2\ep^4 T^{ij} \AND
\nabla_{i} T^{ij} = 0, } where \eqn{EEdefs}{ T^{ij} = (\rho + \ep^2 p)v^i v^j
+ p g^{ij} \AND v^i v_i = -\frac{1}{\ep^2}. }
In this formulation, the fluid four-velocity
$v^i$, the fluid density $\rho$, the fluid pressure $p$, the
metric $g_{ij}$, and the coordinates $(x^i)$ $i=1,\ldots,4$ are
dimensionless.  By assumption, the $(x^i)$ are global Cartesian
coordinates on spacetime $M \cong \Rbb^3\times [0,T)$, where the $(x^I)$
$(I=1,2,3)$ are spatial coordinates that cover $ \Rbb^3$, and
$t=x^4/v_T$ is a \emph{Newtonian time coordinate} that covers the interval
$[0,T)$.
By a choice of units, we can and will set
$v_T=1$.

Post-Newtonian expansions for the Einstein-Euler system refer
to expansions of solutions to this system in the parameter $\ep$, about $\ep =0$, where
the lowest expansion term is governed by the Poisson-Euler equations
of Newtonian gravity:
 \lalign{newtB}{
\del_t \rhoh + \del_I(\rhoh\wh^I) & = 0 \, , \label{newtB.1}\\
\rhoh(\del_t \wh^J + \wh^I\del_I \wh^J) & =
-(\rhoh\del^J\Phih + \del^J\ph) \, , \label{newtB.2} \\
\Delta \Phih &=    \rhoh \, . \label{newtB.3}
 }
Here $\rhoh$, $\ph$, and $\wh{}^J$ are the fluid density, pressure,
and three velocity, respectively.

Formal calculational schemes for determining the
post-Newtonian expansion coefficients and the equations they
satisfy exist, and are in wide use by physicists \cite{Blan,FI}.
In fact, these post-Newtonian computational
schemes are one of the most important techniques in general relativity
for calculating physical quantities for the
purpose of comparing theory with experiment. For example, in gravitational wave
astronomy, post-Newtonian expansions are used to calculate
gravitational wave forms that are emitted during gravitational collapse
\cite{Blan}.

It is important to stress that the formal post-Newtonian expansion
schemes all implicitly
rely on the assumption that the expansions exist and approximate
solutions to general relativity. Therefore, to establish existence
of such approximations, and to answer questions about their range
of validity,
a different approach must be taken to the problem.
In \cite{Oli06}, we took a first step in analyzing this problem
by proving the existence of a wide class of one-parameter families
of solutions to
the Einstein-Euler equations that converged in a suitable
sense to the Poisson-Euler equations in the limit $\ep \searrow 0$.
We also remark that similar results were also established, using a
different method, by Alan Rendall \cite{Ren94}
for the Einstein-Vlasov equations.

In this paper, we use the results of \cite{Oli06} to
prove the existence of a large class of solutions to the Einstein-Euler
equations that can be expanded in $\ep$ to the first post-Newtonian order.
Moreover, we demonstrate the existence of convergent expansions in
$\ep$ for solutions to the Einstein-Euler equations. These expansions
are, in general, not of the post-Newtonian type since the expansion coefficients
can depend on $\ep$. Nevertheless, the expansions are convergent, and
therefore, represent a kind of generalized post-Newtonian expansion.
We note that analogous expansions for the Vlasov-Maxwell equations and Vlasov-Nordst\"{o}m
equations have been rigorously analyzed in  \cite{BauKunz05,Bau05,Bau08}.

The difficulty in analyzing the post-Newtonian expansions arise
from the fact that the limit $\ep \searrow 0$ is singular. To analyze
this limit, we follow the approach of \cite{Oli06}, which
requires that the metric $g_{ij}$ and the fluid velocity
$v^i$ are replaced with new variables that are compatible with the limit
$\ep \searrow 0$. The new gravitational variable is a density $\ufb^{ij}$
defined via the formula
\leqn{metrecA}{
g^{ij} = \frac{\ep}{\sqrt{-\det(Q)}}Q^{ij}
}
where
\leqn{metrecB}{
Q^{ij} = \begin{pmatrix} \delta^{IJ} & 0 \\ 0 & 0 \end{pmatrix}
+  \ep^2 \begin{pmatrix} 4 \ufb^{IJ} & 0 \\ 0 & -1 \end{pmatrix}
+ 4\ep^3 \begin{pmatrix} 0 & \ufb^{I4} \\ \ufb^{J4} & 0 \end{pmatrix}
+ 4 \ep^4\begin{pmatrix} 0 & 0 \\ 0 & \ufb^{44} \end{pmatrix} .
}
From this, it not difficult to see that the density $\ufb^{ij}$ is equivalent to the metric $g_{ij}$
for $\ep > 0$, and is well defined at $\ep =0$.
For the fluid,  a new velocity variable $w^i$ is defined by
\leqn{wdef.intro}{
v^I = w^I \AND w^4=\frac{v^4-1}{\ep}\, .
}

For technical reasons, we assume an isentropic equation of
state
 \leqn{eos}{ p =
K\rho^{(n+1)/n}, } where $K \in \Rbb_{>0}$, $n\in \Nbb$. This
allows us to use
 a technique of Makino \cite{Mak} to regularize the fluid
equations by the use of the fluid density variable
\leqn{dendef}{ \rho = \frac{1}{\bigl(4Kn(n+1)\bigr)^n}\alpha^{2n}. }
The resulting system can be put into
a symmetric hyperbolic system that
is regular across the
fluid-vacuum interface.
In this way,
it is possible to construct solutions to the Einstein-Euler
equations that represent compact gravitating fluid bodies (i.e. stars)
both in the Newtonian and
relativistic setting \cite{Mak,Ren92}. In the Newtonian setting,
this is straightforward to see. Using \eqref{eos}
and \eqref{dendef}, the Poisson-Euler equations \eqref{newtB.1}-
\eqref{newtB.3} imply that
\lalign{newtA}{
\del_t\alphah &= -\wh^I\del_I
\alphah -\frac{\alphah}{2n}\del_I\wh^I, \label{newtA.1} \\
\del_t \wh^J &= -\frac{\alphah}{2n}\del^J\alphah - \wh^I\del_I
\wh^J -\del^J\Phih \label{newtA.2}, \\
\Delta \Phih &=  \rhoh \qquad \bigl(\rhoh :=
(4Kn(n+1))^{-n}\alphah^{2n}\bigr), \label{newtA.3} }
which is readily seen to be regular even across regions where $\alphah$
vanishes.

As discussed by Rendall \cite{Ren92}, the type of fluid solutions
obtained by the Makino method have freely falling boundaries and hence
do not include static stars of finite radius,  and consequently this
method is far from ideal. However, in trying to understand the
post-Newtonian expansions, these solutions
are general enough to obtain a comprehensive
understanding of the mathematical issues involved in the post-Newtonian
expansions.

As in \cite{Oli06}, our approach to the problem of post-Newtonian
expansions is to use the gravitational and matter
variables $\{\ufb^{ij},w^i,\alpha\}$ along with a harmonic
gauge to put the Einstein-Euler equations into
a singular  (non-local) symmetric hyperbolic system of the form
\leqn{EFsym2.intro}{ b^0(\epsilon W)\partial_t W =
\frac{1}{\epsilon}c^I\partial_I W + b^I(\epsilon,W)\partial_I W +
F(\epsilon,W).}
Singular hyperbolic systems of this form have been
extensively studied in the articles
\cite{BK,KM82,Kreiss,Scho86,Scho88}.  Especially relevant for our purposes,
is the paper \cite{Scho88}. There, a systematic procedure for constructing
rigorous expansions to singular symmetric hyperbolic systems is developed
(see also \cite{KM82,Kreiss}).
However, the techniques of \cite{BK,KM82,Kreiss,Scho86,Scho88} cannot be applied directly to our case.
The reason for this is that the initial data for the system
\eqref{EFsym2.intro} must include a $1/r$ piece for the metric and cannot lie in the Sobolev space $H^k$. This problem was overcome in \cite{Oli06}
by using a one parameter family $H^k_{\delta,\ep}$ of weighted Sobolev
spaces that include $1/r$ type fall off for $\ep >0$, and reduce
to the standard Sobolev spaces $H^k$ in the limit $\ep \searrow 0$.
We again use these weighted Sobolev spaces, this time to generalize the
results of \cite{Scho88} so that we can
apply them to the problem of generating rigorous post-Newtonian expansions.

The next theorem is the main result of this paper, and the proof can be found
in section \ref{fpne}. The definition of the spaces $H^k_\delta$, $H^k_{\delta,\ep}$,
and $X_{T,s,k,\delta}$ can be found in Appendices \ref{winq} and \ref{hyp}.
\begin{thm} \label{mth} \mnote{[mth]}
Suppose $-1<\delta < -1/2$, $s\geq 3$, $k\geq 3+s$,
$\alphao,\wo{}^I,\zf_4^{IJ}\in H^k_{\delta-1}$,
$\ff \in H^{k-2}_{\delta-2}$, $\text{\rm supp}\, \alphao\subset B_{R}$,
and let $T^M_0$ is the maximal existence time (see Proposition
\ref{cogA}) for solutions
to the Poisson-Euler-Makino equations \eqref{newtA.1}-\eqref{newtA.3}
with initial
data $\alphah(0)= \alphao$, $w^I(0)=\wo{}^I$. Then for any $T_0 < T$
there exists an $\ep_0 >0$, and maps
\alin{mth1}{
&\ufb_\ep^{ij}(t) \quad : \quad \ufb_\ep^{ij}(t)-\ufb^{ij}(0),\; \del_I \ufb^{ij}_\ep(t),\; \del_t\ufb^{ij}_\ep(t) \in X_{T_0,s,k,\delta-1}\quad 0<\ep \leq \ep_0, \\
&\alpha_\ep(t),\; w^i_\ep(t) \in X_{T_0,s,k,\delta-1}  \quad  0 < \ep \leq \ep_0 ,\\
& \alphah(t),\; \wh^I(t) \in X_{T_0.s,k,\delta-1}, \quad  \Phih(t)\in X_{T_0,s,k+2,\delta}\quad
\text{with} \quad \del_t\Phih(t)\in X_{T_0,s,k+1,\delta-1} ,\\
& \oset{q}{\ufb}{}^{ij}(t) \quad :
\quad \oset{q}{\ufb}{}^{ij}(t)-\oset{q}{\ufb}{}^{ij}(0),\;
\del_I\oset{q}{\ufb}{}^{ij}(t) \in X_{T_0,s-q,k-q,\delta-1} \quad
q=1,2, \\
& \oset{q}{\alpha}(t), \; \oset{q}{w}{}^i(t) \in
X_{T_0,s-q,k-q,\delta-1} \quad q=1,2, \\
& \oset{q}{\ufb}{}_\ep^{ij}(t) \quad : \quad
 \oset{q}{\ufb}{}_\ep^{ij}(t)-\oset{q}{\ufb}{}_\ep^{ij}(0),\;
\del_I\oset{q}{\ufb}{}_\ep^{ij}(t) \in X_{T_0,s-3,k-3,\delta-1}\quad
(q,\ep)\in \mathbb{Z}_{\geq 3}\times (0,\ep_0],\\
& \oset{q}{\alpha}_\ep(t), \; \oset{q}{w}{}_\ep^i(t) \in
X_{T_0,s-3,k-3,\delta-1} \quad (q,\ep)\in \mathbb{Z}_{\geq 3}\times (0,\ep_0], \\
}
such that
\begin{itemize}
\item[(i)] the triple $\{\ufb^{ij}_\ep(t),\alpha_\ep(t),w^i_\ep(t)\}$ determines, via
formulas \eqref{metrecA}-\eqref{dendef},  a solution
to the Einstein-Euler equations \eqref{EEeqn} in the harmonic gauge for $0<\ep\leq \ep_0$ on
the spacetime region $(x^I,t=x^4) \in D=\Rbb^3\times [0,T_0)$,
\item[(ii)] $\del_t\ufb_\ep^{IJ}(0) = \ep^2 \zf_4^{IJ}$, $\del_t^2 \ufb_\ep^{IJ}(0) = \ep^2 \ff^{IJ}$,
$\alpha_\ep(0)=\alphao$, and $w^I_\ep(0) = \wo^{I}$ for $0<\ep \leq \ep_0$,
\item[(iii)] $\{\alphah(t),\wh{}^I(t),\Phih(t)\}$ is the unique solution
to the Poisson-Euler-Makino equations \eqref{newtA.1}-\eqref{newtA.3} with
initial data $\alphah(0)=\alphao$, $\wh{}^I(0)=\wo^I$,
\item[(iv)] for $q=1,2$, $\{\oset{q}{\ufb}{}^{ij}(t),\oset{q}{\alpha}(t),\oset{q}{w}{}^i(t)\}$ satisfies a linear (non-local) symmetric hyperbolic system
that only depends on $\{\alphah(t),\wh{}^I(t),\Phih(t)\}$ if $q=1$,
and $\{\alphah(t),\wh{}^I(t),\Phih(t),\oset{1}{\ufb}{}^{ij}(t),\oset{1}{\alpha}(t),\oset{1}{w}{}^i(t)\}$ if $q=2$,
\item[(v)] for $q\in \Zbb_{\geq 3}$,  $\{\oset{q}{\ufb}{}_\ep^{ij}(t),
\oset{q}{\alpha}_\ep(t),\oset{q}{w}{}_\ep^i(t)\}$ satisfies a linear (non-local) symmetric hyperbolic system
that only depends on $\ep$,  $\{\alphah(t),\wh{}^I(t),\Phih(t)\}$,
 $\{\oset{p}{\ufb}{}^{ij}(t),\oset{p}{\alpha}(t),\oset{p}{w}{}^i(t)\}$
for $p=1,2$, and
 $\{\oset{p}{\ufb}_\ep{}^{ij}(t),\oset{p}{\alpha}_\ep(t),
\oset{p}{w}{}_\ep^i(t)\}$ for $p=3,4,\ldots,q-1$,
\item[(vi)]
 $\{\ufb^{ij},\alpha_\ep(t),w^i_\ep(t)\}$
and $\{\oset{q}{\ufb}{}_\ep^{ij}(t),
\oset{q}{\alpha}_\ep(t),\oset{q}{w}{}_\ep^i(t)\}$ for $q\in \Zbb_{\geq 3}$,
satisfy the following estimates:
\alin{mth2}{
&\norm{\ufb^{ij}_\ep(t)}_{L^2_\delta}
+ \norm{\del_I\ufb_\ep^{ij}(t)}_{H^k} +\ep\norm{\del_t\ufb_\ep^{ij}(t)}_{H^k}
+\ep\norm{\del_t\del_I\ufb_\ep^{ij}(t)}_{H^{k-1}}
+\ep^2\norm{\del_t^2\ufb_\ep^{ij}(t)}_{H^{k-1}} \lesssim 1,\\
& \norm{\alpha_\ep(t)}_{H^{k}}+\norm{w^i_\ep(t)}_{H^{k}}
+ \norm{\del_t\alpha_\ep(t)}_{H^{k-1}} + \norm{\del_t w^i_\ep(t)}_{H^{k-1}} \lesssim 1, \\
 &\norm{\oset{q}{\ufb}{}_\ep^{ij}(t)}_{L^2_\delta}
+ \norm{\del_I\oset{q}{\ufb}{}_\ep^{ij}(t)}_{H^{k-3}} +
\ep\norm{\del_t\oset{q}{\ufb}{}_\ep^{ij}(t)}_{H^{k-3}}
+\ep\norm{\del_t\del_I\oset{q}{\ufb}{}_\ep^{ij}(t)}_{H^{k-4}}
+\ep^2\norm{\del_t^2\oset{q}{\ufb}{}_\ep^{ij}(t)}_{H^{k-4}} \lesssim 1,\\
& \norm{\oset{q}{\alpha}_\ep(t)}_{H^{k-3}}+\norm{\oset{q}{w}{}_\ep^i(t)}_{H^{k-3}}
+ \norm{\del_t\oset{q}{\alpha}_\ep(t)}_{H^{k-4}}+ \norm{\del_t
\oset{q}{w}{}_\ep^i(t)}_{H^{k-4}} \lesssim 1,
}
for all $(t,\ep) \in [0,T_0)\times (0,\ep_0]$, and
\item[(vii)]  $\{\ufb^{ij},\alpha_\ep(t),w^i_\ep(t)\}$ admits
convergent expansions (uniform for $0 < \ep \leq \ep_0$) of the form
\alin{mth3}{
\ufb^{ij}_\ep&= \delta^i_4\delta^j_4\Phih + \sum_{q=1}^2
\ep^q \oset{q}{\ufb}{}^{ij} +
\sum_{q=3}^\infty \ep^q \oset{q}{\ufb}{}_\ep^{ij}, \\
\ep^\nu\del_t^\nu\del_I\ufb^{ij}_\ep &= \ep^\nu
\delta^i_4\delta^j_4\del_t^\nu\del_I\Phih + \sum_{q=1}^2
\ep^{q+\nu} \del_t^\nu\del_I\oset{q}{\ufb}{}^{ij} +
\sum_{q=3}^\infty \ep^{q+\nu} \del_t^\nu\del_I\oset{q}{\ufb}{}_\ep^{ij}
&& \nu=0,1,   \\
\ep^\nu \del_t^\nu \ufb^{ij}_\ep &= \ep^\nu \delta^i_4\delta^j_4
\del_t^\nu \Phih + \sum_{q=1}^2
\ep^{q+\nu} \del_t^\nu \oset{q}{\ufb}{}^{ij} +
\sum_{q=3}^\infty \ep^{q+\nu} \del_t^\nu \oset{q}{\ufb}{}_\ep^{ij}
&& \nu=1,2, \\
\del_t^\nu \alpha_\ep & =
\del_t^\nu \alphah + \sum_{q=1}^2 \ep^q\del_t^\nu \oset{q}{\alpha}
+\sum_{q=3}^\infty \ep^q\del_t^\nu \oset{q}{\alpha}_\ep &&
\nu = 0,1,\\
\del_t^\nu w^i_\ep & =
\del_t^\nu \wh{}^i + \sum_{q=1}^2 \ep^q\del_t^\nu \oset{q}{w}{}^i
+\sum_{q=3}^\infty \ep^q\del_t^\nu \oset{q}{w}{}_\ep^i &&
\nu = 0,1,
}
where the first expansion is convergent in $C^0([0,T_0);L^2_\delta)$,
and the rest are convergent in both
 $C^0([0,T_0);H^{k-4})$ and $C^0([0,T_0);H^{k-4}_{\delta-1,\ep})$.
\end{itemize}
\end{thm}

\begin{rem} {\rm $\;$
\vspace{0.01cm}
\begin{itemize}
\item[(a)] For $q=1,2$, the equations satisfied by $\{\oset{q}{\ufb}{}^{ij},\oset{q}{\alpha},\oset{q}{w}{}^i\}$
are the ones obtained by directly substituting the expansions of Theorem \ref{mth} (vii) into the Einstein-Euler equations
and collecting terms to order $\ep^2$, and therefore coincide with the standard first post-Newtonian expansions.
\item[(b)] The equations satisfied by $\{\oset{q}{\ufb}_\ep{}^{ij}, \oset{q}{\alpha}_\ep, \oset{q}{w}_\ep{}^i\}$
for $q \geq 3$ can be determined from the equations satisfied by the $\oset{q}{W}_\ep$ defined in
the proof of Theorem \ref{hoeA}.
\end{itemize}
}
\end{rem}

To facilitate comparisons of the approach taken in this paper with previous studies, we define the following
$\ep$-independent quantities:
\eqn{hkdef}{
\oset{q}{h}{}^{ij} = \bigl(4\oset{q}{\ufb}{}^{ij} - 2\eta_{k\ell}\oset{q}{\ufb}{}^{k\ell}\eta^{ij}\bigr) \qquad q=1,2,
}
where $(\eta_{ij}) = \diag(1,1,1,-1)$.
Then a straightforward calculation, using statement (vii) of Theorem \ref{mth} and
formulas \eqref{metrecA}-\eqref{metrecB}, shows that the metric $g_{ij}$ can
be expanded as follows
\alin{gexp}{
g_{44} & = -\frac{1}{\ep^2}-2\Phih - \ep \oset{1}{h}{}^{44} - \ep^2
\Bigl(3\bigl(\Phih\bigr)^2 + \oset{2}{h}{}^{44} \Bigr)  + \text{O}(\ep^3),\\
g_{4I} & = \ep^2 \oset{1}{h}{}^{4I} + \ep^3\oset{2}{h}{}^{4I} + \text{O}(\ep^4),
\intertext{and}
g_{IJ} & = \delta_{IJ} -2\ep^2\delta_{IJ}\Phih -\ep^3\oset{1}{h}{}^{IJ} 
-\ep^4\Bigl( \bigl(\Phih\bigr)^2\delta_{IJ}+\oset{2}{h}{}^{IJ} \Bigr) + \text{O}(\ep^5).
}
It is worthwhile to note that higher order expansions in $\ep$ can be generated for the metric $g_{ij}$ using part (vii) of Theorem \ref{mth}. These higher order terms will, in general, depend on $\ep$ in a non-analytic fashion, and therefore, without further analysis, the relation
of these expansion terms to the standard post-Newtonian expansions is not clear.

\sect{EE}{Einstein-Euler equations}

In this section, we quickly review the formulation of the
Einstein-Euler equation used in \cite{Oli06} to analyze the limit
as $\ep \searrow 0$.

\subsect{red}{Reduced Einstein equations}

As discussed in the introduction, we use a symmetric
tensor density $\ufb^{ij}$ instead of the metric $g^{ij}$, which
for $\ep>0$ completely determines the metric via the
formula
\leqn{den2met}{ (g^{ij}) = \frac{1}{\sqrt{|\gb|}}
\begin{pmatrix}
\gfb^{IJ} & \ep \gfb^{I4} \\
\ep \gfb^{4J} & \ep^2 \gfb^{44}
\end{pmatrix},
}
where
\leqn{udensdef}{
\gfb^{ij} := \eta^{ij} + 4\ep^2 \ufb^{ij}\, , \quad |\gb| := - \det(\gfb^{ij}),
}
and \eqn{minkowski}{ \eta^{ij} = \begin{pmatrix}
\id_{\!\!3\times 3} & 0 \\
0 & -1
\end{pmatrix}.
}
To fix the gauge,  we let
\eqn{dbardef}{
\delb_{k} = \left\{ \begin{array}{ll}
\partial_{I} & \text{if $k=I$}\\
\ep \partial_t & \text{if $k=4$}
\end{array} \right. \, ,
}
and demand that \leqn{harm}{
\partialb_{i}\ufb^{ij}
= 0. } For $\ep >0$, this condition is easily seen to
be equivalent to the harmonic gauge
\leqn{harmBa}{ \del_k \gf^{kj}= 0.}
Here
$\gf^{ij} = \sqrt{-\det(g_{k\ell})}g^{ij}$ is the metric density
in the coordinates $(x^i)$.

Next,
defining \lalign{fo}{
\uf^{ij} & := \ep \ufb^{ij}, \label{fo.1} \\
\uf^{ij}_{k} & := \delb_k\ufb^{ij}, \label{fo.2} \\
\ufv^{ij} & := ( \uf_4^{ij}, \uf_J^{ij},
\uf^{ij})^T, \label{fo.3} \\
(\gfb_{ij}) &:= (\gfb^{ij})^{-1} \, , \label{fo.4}\\
A^{ij} & := 2\bigl(\Half \gfb_{k\ell} \gfb_{mn} - \gfb_{km}
\gfb_{\ell n} \bigr) \bigl(\gfb^{ip} \gfb^{jq} - \Half \gfb^{ij}
\gfb^{pq} \bigr)\partialb_{p} \ufb^{k\ell}\partialb_{q}\ufb^{mn}
\label{fo.5},\\
B^{ij} & := 4\gfb_{k\ell}\bigl(2\gfb^{n(i }
\partialb_{m}\ufb^{j)\ell}\partialb_{n}\ufb^{k m} - \Half
\gfb^{ij}\partialb_{m}\ufb^{k n}\partialb_{n}\ufb^{m \ell}
-\gfb^{mn}\partialb_{m}\ufb^{ik}\partialb_{n}\ufb^{j\ell}\bigr)\label{fo.6},
\intertext{and}
C^{ij} & := 4\bigl(\partialb_k\ufb^{ij}\partialb_{\ell}\ufb^{k\ell}-
\partialb_k\ufb^{i\ell}\partialb_\ell\ufb^{jk})\label{fo.7}, \\
} the Einstein equations $G^{ij} = 2\ep^4 T^{ij}$, in the
harmonic gauge, can be written in first order form as \leqn{fo11}{
A^4 (\ep\uf)\partial_t\ufv^{ij} =
\frac{1}{\ep}C^{I}\partial_I\ufv^{ij} +
A^I(\uf)\partial_I\ufv^{ij} + \Fb_0^{ij}(\ufv) + \ep
\Fb_1^{ij}(\ufv,\ep\ufv) -\frac{1}{\ep}(\Tc^{ij},0,0)^T, } where
 \lalign{fo12}{ A^{4}(\ep\uf) &=
\begin{pmatrix}
1-4\ep\uf^{44} & 0 & 0\\
0 & \delta^{IJ}+4\ep\uf^{IJ} & 0 \\
0 & 0 & 1
\end{pmatrix},
\label{fo12.1} \\
 C^{I} &= \begin{pmatrix}
0 & \delta^{IJ} & 0 \\
\delta^{IJ} & 0 & 0 \\
0 & 0 & 0
\end{pmatrix}, \label{fo12.2} \\
 A^{I}(\uf) &= \begin{pmatrix}
4\uf^{4I} & 4\uf^{IJ} & 0 \\
4\uf^{IJ} & 0 & 0 \\
0 & 0 & 0
\end{pmatrix} \, , \label{fo12.3} \\
\Fb_0^{ij}(\ufv) &= (0,0,\uf^{ij}_4)^T
\, , \label{fo12.4} \\
\Fb_1^{ij}(\ufv,\ep \ufv) &= (A^{ij}+B^{ij}+C^{ij},0,0)^T, \label{fo12.5} \intertext{and}
 \frac{1}{\ep}(\Tc^{ij}) &=
\begin{pmatrix}
0 & 0 \\
0 & \ep^{-1}\rho
\end{pmatrix}
+ \Sc^{ij} \label{fo12.6}}
with
\lalign{fo13}{ (\Sc^{ij}) & = \rho
\begin{pmatrix}
0 & |\gb|v^{I}v^{4} \\
|\gb|v^{J}v^{4}  & \ep^{-1}\bigl[(|\gb|-1)(v^{4})^2 +((v^4)^2-1)
\bigr]\end{pmatrix} \notag\\
& \quad + \ep |\gb|\begin{pmatrix} (\rho+\ep^2 p) v^{I}v^{J} +
|\gb|^{-1/2}p(\delta^{IJ}+4\ep\uf^{IJ}) &
\ep pv^{I}v^{4} +  4 \ep |\gb|^{-1/2}p\uf^{I4} \\
\ep pv^{J}v^{4} +  4 \ep |\gb|^{-1/2}p\uf^{4J} & p (v^4)^2 +
|\gb|^{-1/2}p(-1+4\ep\uf^{44})
\end{pmatrix}.\label{fo13.1}
}

Letting \leqn{fluvars}{ \wv = (\alpha,w^i)^T, } we can decompose $\Sc^{ij}$ as
\leqn{fo14}{ \Sc^{ij} = \Sc_0^{ij} + \ep \Sc_1^{ij}, } where
\leqn{fo15}{ \Sc_0^{ij}(\uf,\wv,\ep\uf,\ep \wv)  = \rho
\begin{pmatrix} 0 & |\gb|w^I(1+\ep w^4) \\ |\gb| w^J(1+\ep w^4) &
\ep^{-1}\bigl[  (|\gb|-1)(1+\ep w^4)^2 + \bigl((1+\ep w^4)^2 -1
\bigr) \bigr]
\end{pmatrix},
} and \leqn{fo16}{ \Sc_1^{ij}(\wv,\ep\uf,\ep\wv) =
|\gb|\begin{pmatrix} \rho w^I w^J+p\ep w^I \ep
w^J+|\gb|^{-1/2}p\gfb^{IJ} & p \ep w^I(1+\ep
w^4)+4|\gb|^{-1/2}p\ep\uf^{I4} \\
p \ep w^J(1+\ep w^4)+4|\gb|^{-1/2}p\ep\uf^{J4} & p(1+\ep
w^4)^2+|\gb|^{-1/2} p (-1+4\ep \uf^{44})
\end{pmatrix}. }
 We will
refer to the gauge fixed Einstein equation \eqref{fo11} as the
\emph{reduced Einstein equations}. Because of the matrix inversion
\eqref{fo.5} used to define the inverse density $\gfb_{ij}$, the
reduced Einstein equations will be well defined provided
\eqn{fo17}{ \ep\uf \in \Vc = \{\, (r^{ij})\in \Mbb_{4\times
4} \, | \det(\eta^{ij}+4r^{ij}) > 0\, \}\,. }

\subsect{eul}{Euler equations}

In \cite{Oli06}, we also showed that if we use the fluid variables
\eqref{fluvars},
and choose initial data that satisfies
\leqn{fcon1}{ 0= \Nc := \ep v_i v^i
+1/\ep = \ep \gb_{44}(1/\ep + w^4)^2 +1/\ep + 2\gb_{4J}(1 + \ep
w^4) w^J +\ep\gb_{IJ}w^Iw^J , }
then the Euler equations $\nabla_i T^{ij}=0$ are equivalent to the system
\leqn{eul13}{a^4 \del_4 \wv = a^I\del_I\wv + b,}
where
\lalign{adef}{
\vb^I &= v^I \, , \quad \vb^4 = \frac{v^4}{\ep} , \label{adef.1} \\
\gb^{ij} &= \frac{1}{\sqrt{|\gb|}}\gfb^{ij} \, , \quad (\gb_{ij})=(\gb^{ij})^{-1} , \label{adef.2} \\
 h &= \left( 1 + \frac{1}{4n(n+1)}(\ep\alpha)^2\right)\, , \quad q = \frac{1}{2n h}\alpha\, ,\label{adef.3} \\
L^{i}_{j} &= \delta^i_j + \ep^2 \vb^i\vb_j \, , \quad \vb_j =  \gb_{ij}\vb^i , \label{adef.4}\\
M_{ij} &=  \gb_{ij} + 2\ep^2 \vb_i\vb_j  , \label{adef.5} \\
 \Gammab^{k}_{ij} &= \ep^2\bigl(\gfb^{km}(2\gfb_{i\ell}\gfb_{jp} - \gfb_{ij} \gfb_{\ell
p})\partialb_{m}\ufb^{\ell p} + 2( \gfb_{\ell
p}\delta^{k}_{(i}\partialb_{j)}\ufb^{\ell p}-
2\gfb_{\ell(i}\partialb_{j)}\ufb^{k\ell} )\bigr) , \label{adef.6}\\
a^4 &=
\begin{pmatrix}
h^2(1+\ep w^4) & \ep q L^4_j \\
\ep q L^4_i & M_{ij}(1+\ep w^4)
\end{pmatrix} ,\label{adef.7} \\
a^I &=  \begin{pmatrix}-h^2w^I & -q L^I_j \\
-q L^I_i & -M_{ij}w^I\end{pmatrix} , \label{adef.8} \intertext{and}
b &= \begin{pmatrix} -q L^i_j \Gammab^j_{i\ell}\vb^\ell\\
-M_{ij}\Gammab^j_{k\ell}\vb^k\vb^\ell \end{pmatrix} .
\label{adef.9}
}
We also note that
\lalign{aexp}{ a^4 &= \begin{pmatrix} 1 & 0 \\ 0
& \delta_{ij}\end{pmatrix} + \ah^4(\ep\uf,\ep \wv),
\label{aexp.1}
\\ a^I &= \begin{pmatrix} -w^I & -\frac{\alpha}{2n}\delta^I_j \\
-\frac{\alpha}{2n}\delta^I_i & -\delta_{ij}w^I
\end{pmatrix} +  w^I\ah(\ep\uf,\ep\wv) +
\alpha\ah^I(\ep\uf,\ep\wv) , \label{aexp.2}\intertext{and}
b &= \begin{pmatrix} 0 \\
-\eta^{im}\bigl(2\eta_{4\ell}\eta_{4p}+\eta_{\ell
p}\bigr)\uf^{lp}_m - 2\bigl(\eta_{\ell p}\delta^i_4\uf^{\ell
p}_{4}-2\eta_{\ell 4}\uf^{i\ell}_{4}\bigr)
\end{pmatrix}+\begin{pmatrix}\alpha \bh_1(\ep\uf,\ep\wv)\cdot
\ep\uf_k\\
\bh_2(\ep\uf,\ep\wv)\cdot \uf_k\end{pmatrix}, } where
$\{\ah^4,\ah,\ah^I,\bh_1,\bh_2\}$ are analytic in all
their variables provided that $\ep\uf \in \Vc$, $\{\ah^4,
\ah,\ah^I\}$ are symmetric, and $\ah^4(0,0)=0$,
$\ah^I(0,0) = 0$, $\ah(0,0)= 0$, $\bh_1(0,0)= 0$, and
$\bh_2(0,0)=0$.

\sect{nlim}{Uniform existence and the zeroth order equations}

The combined systems \eqref{fo11} and \eqref{eul13} can be written
as \leqn{EFsym1}{ b^0(\ep V,\ep^2 U)\del_t V =
\frac{1}{\ep}c^I\del_I V + b^I(V,\ep U,\ep V,\ep^2 U)\del_I V +
f_0(V,\ep U,\ep V,\ep^2 U) + \ep f_1(V,\ep U,\ep V,\ep^2 U)+
\frac{1}{\ep}g(V),} where
\lalign{EFsym2}{ U & = ( 0 , 0 ,\ufbo^{ij}, 0, 0 )^T,
\qquad\qquad \ufbo^{ij} = \ufb^{ij}\bigl|_{t=0}, \label{EFsym2.1} \\
V & = ( \uf^{ij}_{4} , \uf^{ij}_{J} , \delta\uf^{ij} , \alpha ,
w^i )^T\, ,  \qquad \qquad \delta\uf^{ij}=
\uf^{ij}-\ep \ufbo^{ij} , \label{EFsym2.2} \\
b^0(\ep V,\ep^2 U) & = \begin{pmatrix} A^4(\ep\uf) & 0 \\
0 & a^4(\ep \uf,\ep \wv)  \end{pmatrix} , \label{EFsym2.3} \\
c^I & = \begin{pmatrix}
C^I & 0 \\
0 & 0
\end{pmatrix}
\label{EYsym2.4},\\
b^I(V,\ep U,\ep V,\ep^2 U) & = \begin{pmatrix}  A^I(\uf) & 0 \\
0 & a^I(\wv,\ep \uf, \ep \wv) \end{pmatrix}  , \label{EFsym2.5}  \\
f_0(V,\ep U,\ep V,\ep^2 U) & = \begin{pmatrix}
\Fb_0^{ij}(\ufv)-\Sc_0^{ij}(\uf,\wv,\ep\uf,\ep \wv) \\
b(\ufv,\wv,\ep\ufv,\ep\wv)
\end{pmatrix}    , \label{EFsym2.6} \\
f_1(V,\ep U,\ep V,\ep^2 U) & = \begin{pmatrix}
\Fb_1^{ij}(\ufv,\ep\ufv)-\Sc_1^{ij}(\wv,\ep\uf,\ep \wv) \\
0
\end{pmatrix} \, ,\label{EFsym2.7} \intertext{and}
g(V) & = ( -\delta^{i}_4\delta^{j}_4 \rho(\alpha),
0,\ldots,0)^T  . \label{EFsym2.8}
}
For initial data, we will often use  the following notation: given a
function $z$ that depends on time $t$, we define \eqn{t0}{
\underset{o}{z} = z|_{t=0}\, .}

In addition to solving these equations, we must also solve constraint equations
on the initial data to get a full solution to the Einstein-Euler equations. Letting
\leqn{Gb}{ \Gc^{ij}=
\gfb^{k\ell}\partialb^2_{k\ell} \ufb^{ij}+
\ep^2\bigl(A^{ij} + B^{ij} + C^{ij}\bigr) +  \gfb^{ij}\partialb^2_{k\ell} \ufb^{k\ell} - 2
\partialb^2_{k\ell}\ufb^{k(i}\gfb^{j)\ell} , }
and defining
\eqn{Condef}{ \quad \Cc^J =
\ep^{-1}(\Gc^{4J}- \Tc^{4J}), \quad  \Cc^4 = \Gc^{44} -
\Tc^{44} , \AND \Hc^j = \delb_{i} \ufb^{ij}, }
the constraint equations
to be solved on the initial hypersurface $
S_0$ $=$ $\{(x^I,0)\,|\, (x^I)\in \Rbb^3 \}$
are:
\lalign{con}{
\Cc^j &= 0 \qquad \text{(gravitational constraint equations),} \label{con.1}\\
\Hc^j & = 0 \qquad \text{(harmonic gauge condition),} \label{con.2}\intertext{and}
\Nc & = 0 \qquad  \text{(fluid velocity normalization).}\label{con.3}
}

To fix a  region on which the system where both the evolution \eqref{EFsym1}
and constraint equations \eqref{con.1}-\eqref{con.3} are well
defined, we note from \eqref{fo12.1}, \eqref{aexp.1}, and the
invertibility of the Lorentz metric $(\eta^{ij})$ that there
exists a constant $K_0>0$ such that \leqn{contA1}{ -\det(\eta^{ij}
+ 4\ep\uf^{ij}) > 1/16\, , \quad 1+\ep w^4
> 1/16\, ,} \leqn{contA2}{ \quad A^4(\ep \uf) \geq \frac{1}{16} \id\, ,
\quad a^4(\ep\uf,\ep \wv) \geq \frac{1}{16}\id, } and
\leqn{contA3}{ |A^4(\ep \uf)|\leq 16 \, , \quad |a^4(\ep\uf,\ep
\wv)| \leq 16} for all $|\ep\uf|\leq 2K_0$, $|\ep
w^i|\leq 2K_0$, $|\ep \alpha| \leq 2K_0$. The choice of the bounds
$1/16$ and $16$ is somewhat arbitrary, and they can be replaced by
any number of the form $1/M$ and $M$ for any $M > 1$ without
changing any of the arguments presented in the following sections.
However, since we are interested in the limit $\ep \searrow 0$, we
lose nothing by assuming $M=16$.

\subsect{idat}{Newtonian initial data}
In \cite{Oli06}, we proved the following theorem, based on previous
work by Lottermoser \cite{Lott}, concerning the existence of
$\ep$-analytic solutions to the constraints \eqref{con.1}-\eqref{con.3}.
Before we state the theorem, we note from \eqref{dendef}, \eqref{eos}, and
the weighted multiplication inequality (see \cite{Oli06} Lemma A.8 ) that if
$\alpha\in H^k_{\delta}$ $(\delta \leq 0,k>3/2)$ then
$\rho,p \in H^k_{\delta}$.
\begin{prop}\label{idatA} \mnote{[idatA]}
Suppose $-1<\delta < 0$, $k > 3/2+1$,  $R>0$
and
$(\tilde{\rho},\tilde{p},\tilde{w}^I,\tilde{\zf}_4^{IJ},\tilde{\zf}^{IJ})
 \in (H^{k-2}_{\delta-2})^2\times H^{k}_{\delta-1}\times
H^{k-1}_{\delta-1}\times B_R(H^{k}_{\delta}) \,.$
Then there exists an
$\ep_0>0$, an open neighborhood $U$ of
$(\tilde{\rho},\tilde{p},\tilde{w}^I,\tilde{\zf}_4^{IJ},\tilde{\zf}^{IJ})$,
and analytic maps $(-\ep_0,\ep_0)\times U \rightarrow
H^{k}_{\delta-1}$ $\; : \;$
$(\ep,\rho,p,w^I,\zf_4^{IJ},\zf^{IJ})$$\mapsto$ $w^4$,
$(-\ep_0,\ep_0)\times U \rightarrow H^{k}_{\delta}$ $\; : \;$
$(\ep,\rho,p,w^I,\zf_4^{IJ},\zf^{IJ})$$\mapsto$ $\phi$,
$(-\ep_0,\ep_0)\times U \rightarrow H^{k}_\delta$ $\; : \;$
$(\ep,\rho,p,w^I,\zf^{IJ}_4,\zf^{IJ})$ $\mapsto$ $\wf^I$ such
that for each $(\rho,p,w^I,\zf^{IJ}_4,\zf^{IJ})\in U$,\\
$(\ep,\rho,p,w^I,w^4,\ufb^{ij}_4,\delb_4\ufb^{ij})$ is a solution to
the three constraints \leqn{idatA2}{ \Cc^j=0\, ,\quad  \Hc^j=0,
\AND \Nc =0,} where \lalign{idatA1}{ (\ufb^{ij}) &
=\begin{pmatrix} \ep \zf^{IJ} &
\ep\wf^I \\
\ep\wf^J &
\phi
\end{pmatrix} \label{idatA1.1},\\
(\del_t\ufb^{ij}) & = \begin{pmatrix} \zf^{IJ}_4 & -
\del_{K}\zf^{KI}
\\ -\del_{K}\zf^{KJ}& -\del_K\wf^K
\end{pmatrix}, \label{idatA1.2} \intertext{and}
w^4 &=  -\frac{1}{\ep} +
\frac{-\ep\gb_{4J}w^J-\sqrt{\ep^2(\gb_{4J}w^J)^2-\gb_{44}
(\ep^2\gb_{IJ}w^Iw^J+1)}}{\ep \gb_{44}}. \label{idatA1.3}}
Moreover, if we let
$\phi_0 = \phi|_{\ep=0}$, $\wf^I_0=\wf^I|_{\ep =0}$,
and $w^4_0 = w^4|_{\ep=0}$,
then $\phi_0$, $\wf^I_0$, and $w^4_0$ satisfy the equations \eqn{idatA3a}{
\Delta\phi_0 = \rho, \quad \Delta\wf^I_0 = -\del_L\zf^{LJ}_4+ \rho w^I, \AND w^4_0 =  0, } respectively.
\end{prop}
In section \ref{hoe}, we show that the analytic dependence of the initial data
on $\ep$ implies that
there exists a corresponding convergent expansion in $\ep$ for the solution generated
from the initial data.

\subsect{eplim}{Uniform existence}

To prove local existence of solutions to
\eqref{EFsym1} on a uniform time interval independent of
$\ep$, we use a
non-local symmetric hyperbolic version of \eqref{EFsym1}. This system
is essentially the one used in \cite{Oli06} to derive
uniform existence, convergence, and error estimates for the limit
$\ep \searrow 0$
of solutions to \eqref{EFsym1}. However, we employ a few refinements
that can be used to simplify the proof in \cite{Oli06}, and will
also be useful
for analyzing the higher order expansions in $\ep$.

Letting $\chi_{\Rb} \in C^\infty_0$ be a cutoff function that satisfies
\eqn{chidef}{ \chi_{\Rb}\bigl|_{B_{\Rb}} = 1, \quad 0\leq \chi_{\Rb} \leq 1\, ,
\AND  \text{supp}\,\chi_{\Rb} \subset B_{2\Rb} \, , }
we replace  $g(V)$ in
\eqref{EFsym1}  with \leqn{newfg}{
g(V)  = ( -\delta^{i}_4\delta^{j}_4 \chi_{\Rb}\rho(\alpha),
0,\ldots,0)^T \,, }
and, following \cite{Oli06}, we define the Newtonian potential by
\leqn{Npot}{
\Delta\Phi =  \chi_{\Rb}\rho \, .
}

Before proceeding, we first recall the following
inequalities from \cite{Oli06}:
\begin{itemize}
\item[(a)] If $\ell>3/2$, there exists a constant $\Csob$ such that
\leqn{sobolev}{
\norm{\cdot}_{L^\infty_{\eta,\ep}} \leq \Csob \norm{\cdot}_{H^\ell_{\eta,\ep}}
\quad \forall \; \ep\in [0,\ep_0]\, .
}
\item[(b)]
For $\ep_0>0$ and $\eta \leq -3/2$,
\leqn{imbed}{
\norm{\cdot}_{H^{\ell}_{\eta,\ep}} \lesssim \norm{\cdot}_{H^\ell_{\eta}}\quad \;
\forall \; \ep\in[0,\ep_0].}
\item[(c)] For $\ep_0 > 0$, and $-2\leq \eta \leq -3/2$,
\leqn{imbedB}{
\norm{\cdot}_{H^\ell}\lesssim \norm{\cdot}_{ H^\ell_{\eta,\ep}} \quad \; \forall \;
\ep \in [0,\ep_0].}
\item[(d)] For $\ep_0>0$ and $\eta \geq -3/2$,
\leqn{imbedC}{
\norm{\cdot}_{L^2_\eta} \lesssim \norm{\cdot}_{L^2_{\eta,\ep}} \quad \;
\forall \; \ep \in [0,\ep_0].}
\item[(d)] If $\ell_2\leq \ell_1$, and $\eta_1 \leq \eta_2$, then
\leqn{imbedD}{
\norm{\cdot}_{H^{\ell_2}_{\ell_2,\ep} } \lesssim \norm{\cdot}_{H^{\ell_1}_{\ell_1,\ep}}.
}
\end{itemize}

\begin{lem} \label{wsysA} \mnote{[wsysA]}
Suppose $\ep_0>0$, $-1<\eta<-1/2$, and $\ell > 3/2$. Then the maps
\eqn{wsysA.1}{\Phi\: : \: H^\ell_{\eta-1,\ep} \longrightarrow
H^{\ell+2}_{\eta} \: : \: \alpha \longmapsto
\Delta^{-1}\bigl( \chi_{\Rb}\rho(\alpha)\bigr)} and
\eqn{wsysA.2}{ \del_I\circ\Phi \: : \: H^{\ell}_{\eta-1,\ep}
\longrightarrow H^{\ell+1}_{\eta-1,\ep} \: : \: \alpha
\longmapsto
\partial_I\Phi(\alpha) }
are uniformly analytic\footnote{See Appendix \ref{winq}
for a definition of the term \emph{uniformly analytic}}
for $\ep \in [0,\ep_0]$.
\end{lem}
\begin{proof}
First we recall that for $-1<\eta<-1/2$,
the Laplacian
\leqn{foeB4}{
\Delta\; : \; H^{\ell+2}_{\eta}
\rightarrow H^{\ell}_{\eta-2}
}
is an isomorphism by
Proposition 2.2 of \cite{Bart86}. Next, by assumption
$\ell > 3/2$, and hence it follows that
the map $H^\ell_{\eta-1,\ep}\ni \alpha
\mapsto \rho= (4Kn(n+1))^{-n}\alpha^{2n}\in H^\ell_{\eta-1,\ep}$ is uniformly
analytic for $\ep \in [0,\ep_0]$ by Lemma \ref{winqE}. Moreover,
the linear map
$H^\ell_{\eta-1,\ep} \ni u \mapsto \chi_{\Rb}u \in H^\ell_{\eta-2}$ is
clearly well defined and uniformly bounded for $\ep \in [0,\ep_0]$.
Since compositions of uniformly analytic maps
are again uniformly analytic, we see that
the map  $H^\ell_{\eta-1,\ep} \ni \alpha \rightarrow
\Delta^{-1}\bigl( \chi_{\Rb}\rho(\alpha)\bigr)\in H^{\ell+2}_{\eta}$ is
uniformly analytic of $\ep \in [0,\ep_0]$.

Next, we recall that differentiation
$H^{\ell+2}_{\eta} \ni u \mapsto \del_Iu \in H^{\ell+1}_{\eta-1}$ is a bounded
linear map, and the imbedding
$H^{\ell+1}_{\eta-1} \subset H^{\ell+1}_{\eta-1,\ep}$
is well defined and uniformly bounded for
$\ep \in [0,\ep_0]$  by \eqref{imbed}. Again using the fact
that uniform analyticity
is preserved under compositions, we get that the
map $\del_I\circ\Phi \: : \: H^{\ell}_{\eta-1,\ep}
\rightarrow H^{\ell+1}_{\eta-1,\ep}$ is uniformly analytic
for $\ep \in [0,\ep_0]$.
\end{proof}
Following \cite{Oli06}, we use the Newtonian potential to define a new
combined gravitational-matter variable $W$ via the formula
\leqn{Wdef}{ W = V - d \Phi,} where \leqn{dPhi}{ d\Phi :=
(0,\delta^i_4\delta^j_4\del_J\Phi(\alpha),0,0,0)\, .}
Notice that the transformation \eqref{Wdef} leaves the matter variables
unaffected. Consequently,  we can define $W$ by \eqn{Wdefa}{ W =
(\uf_4^{ij},W_I^{ij},\delta\uf^{ij},\alpha,w^i)^T, } and treat
$\Phi$ or $d\Phi$ as a function of $W$. In fact, by Lemma
\ref{wsysA}, \leqn{dPhi1}{ H^\ell_{\delta-1,\ep} \ni W \longmapsto
d\Phi  \in H^\ell_{\delta-1,\ep} } defines a uniformly analytic map for
$\ep\in [0,\ep_0]$.

To formulate the
evolution equation entirely in terms of $W$, we need the
``time derivative'' of the $\Phi$ map. So we define \leqn{Phid}{
\Phid(W,\ep U,\ep W, \ep^2 U) :=
\Delta^{-1}\left(\frac{2n \chi_{\Rb}\alpha^{2n-1}}{(4Kn(n+1))^n}\Pi\bigl(a^4(\ep\uf,\ep\wv)^{-1}
\bigl[a^I(\wv,\ep\uf,\ep\wv)\del_I \wv +
b(\ufv,\wv,\ep\ufv,\ep\wv) \bigr] \bigr) \right) } where
$\Pi((\alpha,w^i)^T) = \alpha$ is a constant projection map.
By construction,  $\Phid = \partial_t\Phi$ when evaluated on a solution of
the reduced Einstein-Euler equations.

\begin{lem}\label{wsysB} \mnote{[wsysB]}
Suppose $R_1>0$, $\ep_0>0$, $-1<\eta<1/2$, and $\ell > 3/2$. Then there
exists an $R_2>0$ such that the maps \eqn{wsysB.1}{\Phid\: : \:
B_{R_1}(H^\ell_{\eta-1,\ep})\times
B_{R_2}(H^\ell_{\eta})\times
B_{R_2}(H^\ell_{\eta-1,\ep})\times B_{R_2}(H^\ell_{\eta})
\longrightarrow H^{\ell+1}_{\eta} \: : \:
(W,U,\tilde{W},\tilde{U}) \longmapsto
\Phid(W,U,\tilde{W},\tilde{U}) } and \eqn{wsysB.2}{
\del_I\circ\Phid \: : \: B_{R_1}(H^\ell_{\eta-1,\ep})\times
B_{R_2}(H^\ell_{\eta})\times
B_{R_2}(H^\ell_{\eta-1,\ep})\times B_{R_2}(H^\ell_{\eta})
\longrightarrow H^{\ell}_{\eta-1,\ep} \: : \:
(W,U,\tilde{W},\tilde{U}) \longmapsto
\del_I\bigl(\Phid(W,U,\tilde{W},\tilde{U})\bigr) } are uniformly
analytic
for $\ep \in [0,\ep_0]$.
\end{lem}
\begin{proof}
Fixing $R_1>0$, $\ep_0>0$, $-1<\eta<-1/2$ and
$\ell > 3/2$, it follows directly from Lemmas
\ref{winqCb} and \ref{winqE} that there exists a $R_2>0$ such that
the map
\alin{wsysB.3}{
B_{R_1}(H^\ell_{\eta-1,\ep})\times
B_{R_2}&(H^\ell_{\eta})\times
B_{R_2}(H^\ell_{\eta-1,\ep})\times B_{R_2}(H^\ell_{\eta}) \ni
(W,U,\ep W,\ep^2 U) \longmapsto
 \\
&\chi_{\Rb}\alpha^{2n-1}\Pi\bigl(a^4(\ep\uf,\ep\wv)^{-1}
\bigl[a^I(\wv,\ep\uf,\ep\wv)\del_I \wv +
b(\ufv,\wv,\ep\ufv,\ep\wv) \bigr]
\in H^{\ell-1}_{\eta-2}
}
is uniformly analytic for $\ep\in [0,\ep_0]$. The rest
of the proof now follows from the same arguments used
in the proof of Lemma \ref{wsysA}.
\end{proof}
To fit with the above notation, we define
\eqn{dPhid}{
d\Phid = (0,\delta^i_4\delta^j_4\del_I\Phid,0,0,0)^T \, .
}
Noting that \leqn{shift}{
b^0(\ep V,\ep^2 U) = b^0(\ep W,\ep^2 U) \AND b^I(V,\ep U,\ep V, \ep^2) =
b^I(W,\ep U,\ep W,\ep^2 U), }
we write \eqref{EFsym1} as
\leqn{wsysdef}{
b^0(\ep W,\ep^2 U)\del_t W =
\frac{1}{\ep}c^I\del_I W + b^I(W,\ep U,\ep W,\ep^2 U)\del_I W +
\Fc_0(W,\ep U,\ep W,\ep^2 U) + \ep \Fc_1(W,\ep U,\ep W,\ep^2 U),
}
where
\lalign{Fcaldef}{
\Fc_0(W,\ep U,\ep W,\ep^2 U) &= f_0(W+d\Phi(W),\ep U,\ep(W+d\Phi(W)),\ep^2 U) \notag \\
& \quad -b^0(\ep W,\ep^2 W)d\Phid(W,\ep U,\ep W,\ep^2 U)+ b^I(W,\ep U,\ep W)\del_I d\Phi(W)\,
\label{Fcaldef.1}
 \intertext{and}
\Fc_1(W,\ep U,\ep W,\ep^2 U)) &= f_1(W+d\Phi(W),\ep U,\ep(W+d\Phi(W)),\ep^2 U) \, . \label{Fcaldef.2}
}

\begin{prop} \label{eplimA} \mnote{[eplimA]}
Suppose $-1 < \delta < -1/2$, $\ep_0>0$, $s\in \mathbb{N}_{0}$, $R>0$,
$K_1 < K_0/(2\sqrt{\ep_0}\Csob)$, $\tau \geq 2K_1/\Csob$,
$\Rb > 16\tau+ R$,  $k\geq 3+s$, $\alphao,\wo^I \in H^k_{\delta-1}$,
$\supp\, \alphao \subset B_R$, $\mathfrak{z}^{IJ}\in H^{k+1}_\delta$,
$\mathfrak{z}^{IJ}_4 \in H^k_{\delta-1}$. Let
$\ufbo_\ep^{ij}$, $\del_t\ufbo^{ij}_\ep$ and $\wo^4_\ep$ be the
initial data constructed in Proposition \ref{idatA}, which,
by choosing $\ep_0\leq 1$ small enough, satisfies
\eqn{eplimA.1}{
\Bigl\|\Bigl(\ep\del_t\ufbo^{ij}_\ep,\del_I\ufbo^{ij}_\ep-\delta^i_4
\delta^j_4\del_I\Delta^{-1} \rhoo,0,\alphao,\wo^i_\ep\Bigr)^T\Bigr\|_{H^{k}_{\delta-1,\ep}}\leq K_1\, ,
\AND
\norm{\ufbo^{ij}_\ep}_{H^{k+1}_{\delta}}\leq \frac{K_0}{\sqrt{\ep_0}\Csob}
}for all $\ep \in (0,\ep_0]$.
Then there exists a $T>0$ independent of $\ep \in (0,\ep_0]$,
and maps
\eqn{eplimA.2}{
W_\ep = \bigl(\uf^{ij}_{4,\ep},W^{ij}_{I,\ep},\delta\uf^{ij}_\ep,
\alpha_\ep,w^i_\ep \bigr)^T \in X_{T_\ep,s,k,\delta-1} \qquad 0<\ep \leq \ep_0
}
such that
\begin{itemize}
\item[(i)] $T_\ep \geq T$ for $0\leq \ep \leq \ep_0$,
\item[(ii)] $W_\ep$ is the unique solution to
\eqref{wsysdef} with initial data
\eqn{eplimA.3}{
W_\ep(0) = \Bigl(\ep\del_t\ufbo^{ij}_\ep,\del_I\ufbo^{ij}-\delta^i_4
\delta^j_4\del_I\Delta^{-1} \rhoo,0,\alphao,\wo^i_\ep\Bigr)^T,
}
\item[(iii)]
\gath{epliA.5}{
\norm{W_\ep(t)}_{H^{k}_{\delta-1,\ep}} \leq 2K_1,
\quad
\norm{\del_t W_{\ep}(t)}_{H^{k-1}_{\delta-1,\ep}} \lesssim 1,
\intertext{and}
\max\{\norm{\ep\ufb^{ij}_\ep(t)}_{L^\infty},
\norm{\ep\alpha_\ep(t)}_{L^\infty},\norm{\ep w^i(t)}_{L^\infty}\} < 2K_0
}
for all $(t,\ep) \in [0,T]\times (0,\ep_0]$,
\item[(iv)] if
\gath{eplimA.5a}{
\limsup_{t\nearrow T_\ep}\norm{W_\ep(t)}_{W^{1,\infty}} <\infty \, ,
\intertext{and}
\sup_{0\leq t < T_\ep}\{\norm{\ep\ufb^{ij}_\ep(t)}_{L^\infty},
\norm{\ep\alpha_\ep(t)}_{L^\infty},\norm{\ep w^i(t)}_{L^\infty}\} < 2K_0 \, ,
}
then the solution $W_\ep(t)$ can be uniquely extended for some time $T^*_\ep>T_\ep$,

\item[(v)]
for any time $\tilde{T}_\ep$ which is strictly less than
the maximal existence time and for which
\eqn{eplimA.4a}{
\sup_{0\leq t \leq T_\ep}\{\norm{\ep\ufb^{ij}_\ep(t)}_{L^\infty},
\norm{\ep\alpha_\ep(t)}_{L^\infty},\norm{\ep w^i(t)}_{L^\infty}\} < 2K_0}
holds,
the support of $\alpha_\ep$ satisfies
\eqn{eplimA.4ab}{
\supp\, \alpha_\ep(t) \subset B_{\Rb_\ep} \quad \forall\; t \in [0,\tilde{T}_\ep],
}
where $\Rb_\ep := 16\sup_{0\leq t\leq \tilde{T}_\ep}\norm{w^I_\ep(t)}_{L^\infty}+R$,
\item[(vi)]$\supp\, \alpha_\ep(t) \subset B_{\Rb}$ for all
$(t,\ep) \in [0,T]\times (0,\ep_0]$,
\item[(vii)]$\del_t \ufb^{ij}_{\ep} = \ep^{-1} \ufb^{ij}_{4,\ep}$,
 and $\del_I \ufb^{ij}_{\ep} = W_{I,\ep}^{ij} +
\delta^{i}_4\delta^{j}_4\del_I\Phi(\alpha_\ep)$,
where $\ufb^{ij}_\ep = \ufbo^{ij}_\ep + \ep^{-1}\delta \uf^{ij}$,
\item[(viii)] the triple $\{\ufb^{ij}_\ep,\alpha_\ep,w^i_\ep\}$ determines,
via the formulas \eqref{wdef.intro}, \eqref{dendef}, \eqref{den2met},
and \eqref{udensdef},
a solutions to the full Einstein-Euler system \eqref{EEeqn} in
the harmonic gauge \eqref{harmBa}
on the spacetime region $D_\ep = \Rbb^3\times [0,T]$, and
\item[(ix)] the conclusions (vii)-(viii) continue to hold
on any region of the form $D_\ep = \Rbb^3\times [0,\tilde{T}_\ep]$ provided
$\supp\, \alpha_\ep(t) \subset B_{\Rb}$ for all $0\leq t\leq \tilde{T}_\ep$.
\end{itemize}
\end{prop}
\begin{proof}
\noindent {\bf (i)}-{\bf (iv)}: Given the initial data satisfying
\eqn{eplimA.1a}{
\Bigl\|\Bigl(\ep\del_t\ufbo^{ij}_\ep,\del_I\ufbo^{ij}_\ep-\delta^i_4
\delta^j_4\del_I\Delta^{-1} \rhoo,0,\alphao,\wo^i_\ep\Bigr)^T\Bigr\|_{H^k_{\delta-1,\ep}}\leq K_1\, ,
\AND
\norm{\ufbo^{ij}_\ep}_{H^{k+1}_{\delta}}\leq \frac{K_0}{\sqrt{\ep_0}\Csob}
}for all $\ep \in (0,\ep_0]$, it is not difficult using the
inequalities \eqref{sobolev} and \eqref{imbed},
and
Lemmas \ref{wsysA}, \ref{wsysB},
and \ref{winqE} to verify that
$\norm{W_\ep(0)}_{H^{k}_{\delta-1,\ep}} \leq K_1$, $\norm{\del_tW_\ep(0)}_{H^{k-1}_{\delta-1,\ep}} \lesssim 1$,
and
the evolution equation \eqref{wsysdef} satisfies the conditions
\eqref{Fcond1}-\eqref{bb}. Therefore, it follows directly from
Theorem \ref{hypA} that there exists a time $T>0$ independent of
$\ep \in (0,\ep_0]$ such that
$\norm{W_\ep(t)}_{H^{k}_{\delta-1,\ep}} \leq 2K_1 < 2K_0/(\sqrt{\ep_0}\Csob)$,
and $\norm{\del_t W_\ep(t)}_{H^{k-1}_{\delta-1,\ep}}\lesssim 1$ for all
$0\leq t \leq T$. This proves (i)-(iii).
Statement (iv) also follows directly from Theorem \ref{hypA}.
\bigskip

\noindent {\bf (v)}-{\bf (vi)}: Statement (v) follows
from a slight modification of Lemma 7.2 in \cite{Oli06} while
(vi) follows directly from (iii) and (v).
\bigskip

\noindent {\bf(vii)}-{\bf(ix)}: By (vi) we see that
$V_\ep(t) = W_\ep(t)+d\Phi(W_\ep(t))$ satisfies \eqref{EFsym1} for
$(t,\ep) \in [0,T]\times (0,\ep_0]$. Then the same arguments
used to prove (ii) and (iii) of Proposition 6.1 in \cite{Oli06} can
be employed to
prove the statements (vii)-(ix) of this Proposition.
\end{proof}

\subsect{zeqn}{Zeroth order equation}

In order to discuss equations satisfied by the zeroth and higher
order expansions, we will first introduce some notation.
To begin, we define
\gath{UWvec}{
\ovec{p}{U} = (\oset{0}{U},\oset{1}{U},\ldots,\oset{p}{U} )\, , \quad
\ovec{p}{W} = (\oset{0}{W},\oset{1}{W},\ldots,\oset{p}{W}) \, , \\
\ovec{p}{X} = (\oset{0}{X},\oset{1}{X},\ldots,\oset{p}{X}) \, , \quad
\ovec{p}{Y} = (\oset{0}{Y_I},\oset{1}{Y_I},\ldots,\oset{p}{Y_I}) \, ,
}
and let
\gath{FBdef}{
\Fc_\ep(U,W) = \Fc_0(W,\ep U,\ep W,\ep^2 U) + \Fc_1(W,\ep U, \ep W, \ep^2 U)\, , \\
B_\ep(U,W,Y) = b^I(W,\ep U,\ep W,\ep^2 U)Y_I \, ,
\intertext{and}
B^0_\ep(U,W,X) = b^0(\ep^2 U,\ep W)X \, .
}
\begin{prop} \label{eplimB} \mnote{[eplimB]}
Suppose $\ell>3/2$, $R > 0$, $-1<\eta < -1/2$. Then there exists an $\ep_0 > 0$
such that the maps
\gath{eplimB1}{
\Fc_\ep : B_R(H^\ell_\eta) \times B_R(H^\ell_{\eta-1,\ep}) \longrightarrow
H^\ell_{\eta-1,\ep}, \\
B_\ep : B_R(H^\ell_\eta) \times B_R(H^\ell_{\eta-1,\ep})\times
B_R(H^\ell_{\eta-1,\ep}) \longrightarrow H^\ell_{\eta-1,\ep},
\intertext{and}
B^0_\ep : B_R(H^\ell_\eta) \times B_R(H^\ell_{\eta-1,\ep})\times
B_R(H^\ell_{\delta-1,\ep}) \longrightarrow H^\ell_{\eta-1,\ep}
}
are uniformly analytic for $\ep \in [0,\ep_0]$.
\end{prop}
\begin{proof}
The proof follows directly from Lemmas \ref{wsysA}, \ref{wsysB}, \ref{winqE}, and
the fact that compositions of uniformly analytic functions are
again analytic.
\end{proof}

Next, we define
\gath{DFBdef}{
\oset{p}{\Fc}(\ovec{p-1}{U},\ovec{p}{W}) = \frac{1}{p!}\frac{d^p}{d\ep^p}\Bigl|_{\ep=0} \Fc_\ep (U(\ep),W(\ep), \\
\oset{p}{B}(\ovec{p-1}{U},\ovec{p}{W},\ovec{p}{Y})
= \frac{1}{p!}\frac{d^p}{d\ep^p}\Bigl|_{\ep=0}B_\ep(U(\ep),W(\ep),Y(\ep)),
\intertext{and}
\oset{p}{B}{}(\ovec{p-2}{U},\ovec{p-1}{W},\ovec{p}{X})
= \frac{1}{p!}\frac{d^p}{d\ep^p}\Bigl|_{\ep=0}B^0_\ep(U(\ep),W(\ep),X(\ep)),
}
where
\eqn{UWYepdef}{
U(\ep) = \sum_{q=0}^{p-1} \ep^q \oset{q}{U}\, , \quad
 W(\ep) = \sum_{q=0}^{p-1} \ep^q \oset{q}{W} \, ,
\quad X(\ep)=\sum_{q=0}^{p}\ep^q \oset{q}{X}\, , \AND
 Y(\ep) = \sum_{q=0}^{p} \ep^q \oset{q}{Y} \, .
}

\begin{prop} \label{eplimC} \mnote{[eplimC]}
Suppose $\ell > 3/2$, $R > 0$, $-1 <\eta < -1/2$. Then there exists an $\ep_0 > 0$
such that the maps
\gath{eplimC1}{
\oset{p}{\Fc} : \Bigl(B_R(H^\ell_\eta)\times (H^\ell_\eta)^{p-2}\Bigr)
\times \Bigl(B_R(H^\ell_{\eta-1,\ep})\times(H^\ell_{\eta-1,\ep})^{p-1} \Bigr)
\longrightarrow H^\ell_{\eta-1,\ep} \, , \\
\oset{p}{B} : \Bigl(B_R(H^\ell_\eta) \times (H^\ell_\eta)^{p-2} \Bigr)
\times \Bigl(B_R(H^\ell_{\eta-1,\ep}) \times
(H^\ell_{\eta-1,\ep})^{p-1}\Bigr)\times (H^\ell_{\eta-1,\ep})^{p} \longrightarrow H^\ell_{\eta-1,\ep} \, ,
\intertext{and}
\oset{p}{B^0} : \Bigl(B_R(H^\ell_\eta) \times (H^\ell_\eta)^{p-3} \Bigr)
\times \Bigl(B_R(H^\ell_{\eta-1,\ep}) \times
(H^\ell_{\eta-1,\ep})^{p-2}\Bigr)\times (H^\ell_{\eta-1,\ep})^{p} \longrightarrow H^\ell_{\eta-1,\ep}
}
are uniformly analytic for $\ep \in [0,\ep_0]$. Moreover,
there exists uniformly analytic maps
\gath{eplimC1A}{
\oset{p}{\Fc}{}_{\Rc,\ep} : \Bigl(B_R(H^\ell_\eta)\times (H^\ell_\eta)^{p-2}\Bigr)
\times \Bigl(B_R(H^\ell_{\eta-1,\ep})\times(H^\ell_{\eta-1,\ep})^{p-1} \Bigr)
\longrightarrow H^\ell_{\eta-1,\ep} \, , \\
\oset{p}{B}{}_{\Rc,\ep} : \Bigl(B_R(H^\ell_\eta) \times (H^\ell_\eta)^{p-2} \Bigr)
\times \Bigl(B_R(H^\ell_{\eta-1,\ep}) \times
(H^\ell_{\eta-1,\ep})^{p-1}\Bigr)\times (H^\ell_{\eta-1,\ep})^{p} \longrightarrow H^\ell_{\eta-1,\ep} \, ,
\intertext{and}
\oset{p}{B}{}^{0}{}_{\Rc,\ep} : \Bigl(B_R(H^\ell_\eta) \times (H^\ell_\eta)^{p-3} \Bigr)
\times \Bigl(B_R(H^\ell_{\eta-1,\ep}) \times
(H^\ell_{\eta-1,\ep})^{p-2}\Bigr)\times (H^\ell_{\eta-1,\ep})^{p} \longrightarrow H^\ell_{\eta-1,\ep}
}
that are linear in the variables $\oset{1}{U},\ldots,\oset{p-1}{U}$, $\oset{1}{W},\ldots,\oset{p}{W}$,
$\oset{0}{X},\ldots,\oset{p}{X}$, $\oset{0}{Y},\ldots,\oset{p}{Y}$, and
\gath{eplimC2}{
\frac{1}{\ep^{p+1}}\left[\Fc_\ep(U(\ep),W(\ep))-\sum_{q=0}^p \ep^q \oset{q}{F}
(\ovec{q-1}{U},\ovec{q}{W}) \right] =  \oset{p}{\Fc}{}_{\Rc,\ep}(\ovec{p-1}{U},\ovec{p}{W}) ,\\
\frac{1}{\ep^{p+1}}\left[ B_\ep(U(\ep),W(\ep),Y(\ep))-\sum_{q=0}^p \ep^q \oset{q}{B}
(\ovec{q-1}{U},\ovec{q}{W},\ovec{q}{Y}) \right] = \oset{p}{B}{}_{\Rc,\ep}(\ep,\ovec{p-1}{U},\ovec{p}{W},
\ovec{p}{Y}) ,
\intertext{and}
\frac{1}{\ep^{p+1}}\left[ B^0_\ep(U(\ep),W(\ep),X(\ep))-\sum_{q=0}^p \ep^q \oset{q}{B^0}
(\ovec{q-2}{U},\ovec{q-1}{W},\ovec{q}{X}) \right]
= \oset{p}{B^0}{}_{\Rc,\ep}(\ep,\ovec{p-2}{U},\ovec{p-1}{W},\ovec{p}{X}) .
}
\end{prop}
\begin{proof} The proof follows immediately from the Taylor expansions for
$\Fc_\ep$, $B_\ep$, and $B^0_\ep$ which are uniformly analytic by Proposition \ref{eplimB}.
\end{proof}
We note that from the definition of the above maps, it is clear that
\leqn{Btdef}{
\oset{p}{B} = \oset{0}{b^I}(\oset{0}{W})\oset{p}{Y}_I + \oset{p}{\Bt}(\ovec{p-1}{U},\ovec{p}{W},\ovec{p-1}{Y})
\AND
\oset{p}{B^0} = \oset{p}{X} + \oset{p}{\Bt^0}(\ovec{p-2}{U},\ovec{p-1}{W},\ovec{p-1}{X}),
}
where
\leqn{Btdef2}{
\oset{0}{b^I}(\oset{0}{W}) := b^I(\oset{0}{W},0,0,0) \AND \oset{0}{\Bt} = \oset{0}{\Bt^0} = 0 \, .
}

With our notation fixed, we are now ready to define the zeroth order equations:
\lalign{zerotheqns}{
\del_t \oset{0}{W} &= \oset{0}{b^I}(\oset{0}{W})\del_I \oset{0}{W} + \oset{0}{\Fc}(\oset{0}{W}) +c^I\del_I
\oset{1}{\omega} \label{zerotheqns.1}\, , \\
c^I\del_I \oset{0}{W} & = 0 \label{zerotheqns.2} \, ,\\
\oset{0}{W}(0) & = W_\ep(0)\bigl|_{\ep=0} \label{zerotheqns.3} \, .
}
We showed in \cite{Oli06} that these equation are equivalent to the Poisson-Euler equations
of Newtonian gravity.
To see this, we first note that the Poisson-Euler-Makino system
\eqref{newtA.1}-\eqref{newtA.3} is  (non-local) symmetric hyperbolic,
and thus we can
use the results of Appendix \ref{hyp} to obtain local
existence of solutions.
\begin{prop} \label{cogA} \mnote{[cogA]}
Let $k$, $s$, $\delta$, $\alphao$, and $\wo$ be as in Proposition \ref{eplimA}.
Then there exists a maximal time $T_0^M>0$ and a unique solution \gath{cogA1}{ \alphah, \wh^I
\in C^0([0,T^M_0),H^k_{\delta-1})\cap
C^{1}([0,T_0),H^{k-1}_{\delta-1})\, , \\
\Phih \in C^0([0,T^M_0),H^{k+2}_{\delta})\cap
C^{1}([0,T^M_0),H^{k+1}_{\delta})\, , \quad \del_t\Phih \in
C^{0}([0,T^M_0),H^{k+1}_{\delta-1}) } to
\eqref{newtA.1}-\eqref{newtA.3} satisfying $\alphah(0)=\alphao$ and  $\wh{}^I(0)=\wo^I$. Moreover,
\eqn{cogA2a}{\alphah,
\wh^I \in X_{T^M_0,s,k,\delta-1} \, , \quad
\Phih \in X_{T^M_0,s,k+2,\delta} \, , \quad \del_t\Phih =-\del_I\Delta^{-1}
(\rhoh \wh^I) \in X_{T^M_0,s,k+1,\delta-1} \, , }
and
\eqn{cogA2}{
\supp\, \alphah(t) \subset B_{R(t)} \quad \forall \, t\in [0,T^M_0) \, ,
}
where $R(t)=R + t\sup_{0\leq s \leq t}\,\norm{\wh{}^I(s)}_{L^\infty}$.
\end{prop}
\begin{proof}
From the weighted calculus inequalities of Appendix \ref{winq} (see also
Appendix A of \cite{Oli06}), the Poisson-Euler-Makino system \eqref{newtA.1}-\eqref{newtA.3} satisfies
the conditions required by Theorem \ref{hypA}. Therefore all of the statements
except for the estimate on the support of $\alphah(t)$ follow from this theorem.
To prove the estimate on the support, we note that
$\wh{}^I\in C^1([0,T^M_0),C^1_b(\Rbb))$
by the Sobolev inequality \eqref{sobolev}. Therefore we
can integrate the differential equation $dx^I/dt = \wh^I(t,x)$ to
get a $C^1$ flow $\psi^I_t(x)$ that is defined for all  $(t,x)
\in [0,T_0)\times \Rbb^3$ and satisfies $\psi_0=\id_{\!\!\Rbb^3}$.
For each $x\in \Rbb^3$, define $\alphah^x(t) = \alphah(t,\psi_t(x))$.
The evolution equation \eqref{newtA.1} implies that
\eqn{cogA3}{
\frac{d\;}{dt}\alphah^x(t) + \frac{1}{2n}\del_I\wh^I(x,\psi_t(x))
\alphah^x(t) = 0 \, .
}
By assumption, $\alphah^x(0)= \alphah(0,x) = 0$
for all $x\in E_R:= \Rbb^3\setminus B_R$, and thus
\leqn{cogA4}{
\alphah{}^x(t)=\alphah(t,\psi_t(x)) = 0 \quad \text{for
all $(t,x) \in [0,T^M_0)\times E_R$}
}
by the above differential equation.
Moreover,
\eqn{cogA5}{
|\psi_t(x)-x|\leq \int_{0}^t|\del_s\psi_s(x)|\,ds \leq
\int_0^t|\wh^I(x,\psi_s(x))|\,ds\leq t \sup_{0\leq s\leq t}\norm{\wh^I(s)}_{L^\infty} \, ,
}
and hence it follows from \eqref{cogA4} that
$\text{supp}\, \alphah(t)$ $\subset B_{R(t)}$, where
$R(t)$ $=$ $R$ $+$  $t\, \sup_{0\leq s\leq t}\norm{\wh^I(s)}_{L^\infty}$.
\end{proof}
Using this local existence theorem, the next proposition follows
by straightforward computation.
\begin{prop} \label{eplimD} \mnote{[eplimD]}
Let $\{\alpha(t),\wh^I(t),\Phih(t)\}$ be the solution to
the Makino-Euler-Poisson equations\eqref{newtA.1}-\eqref{newtA.3}
from Proposition \ref{cogA},
and define
\eqn{W0def}{
\oset{0}{W}(t) = (0,-\delta^i_4\delta^j_4\Phih(t),0,\alphah(t),\delta^i_I\wh^I(t))^T
\in X_{T^M_0,s,k,\delta-1}, \AND
\oset{1}{\omega}(t) = (\oset{1}{\omega}_4{}^{ij}(t),
\oset{1}{\omega}_I{}^{ij}(t),0,0,0)^T\, ,
}
where
\eqn{omega1defA}{
\oset{1}{\omega}_4{}^{ij}=\delta^i_4\delta^j_4\del_t\Phih \in X_{T^M_0,s,k+1,\delta-1}
\AND
\oset{1}{\omega}_I{}^{ij} = \del_I\Delta^{-1}\left(2\rhoh\delta_J^{(i}\delta^{j)}_4 \wh^J \right)
\in X_{T^M_0,s,k+1,\delta-1}
\, .
}
Then $\{\oset{0}{W}(t),\oset{1}{\omega}(t)\}$ defines a unique solution to
the
initial value problem
\eqref{zerotheqns.1}-\eqref{zerotheqns.3} on the time interval $0\leq t < T^M_0$.
\end{prop}

\sect{foe}{First order expansion}

By Proposition \ref{idatA}, the initial data $\ufbo_\ep^{ij}$
is analytic in $\ep$ and there exists a convergent
expansion in $H^{k+1}_\delta$ for $\ufbo_\ep^{ij}$ of the form
$\ufbo_\ep^{ij} = \sum_{q=0}^{\infty} \ep^q \oset{q}{\ufbo}{}^{ij}$
for $0\leq \ep \leq \ep_0$.
Consequently, $U$ can be expanded
as $U = \sum_{q=0}^{\infty} \ep^q \oset{q}{U}$, where
$\oset{q}{U}= (0,0,\oset{q}{\ufbo}{}^{ij},0,0)^T$.
Moreover, by Lemma \ref{wsysA} and the inequality \eqref{imbed}, we can expand $W_\ep(0)$
as
\leqn{Wexp}{
W_\ep(0) = \sum_{q=0}^\infty \ep^q \oset{q}{\uset{0}{W}}
}
with the sum converging in $H^{k}_{\delta-1,\ep}$ uniformly for
$0\leq \ep \leq \ep_0$.

We define the second order remainder $\oset{2}{Z}_\ep$ by
\leqn{Z2eqn}{
W_\ep = \oset{0}{W} + \ep(\oset{1}{\omega} + \oset{1}{W}_\ep) + \ep^2\oset{2}{Z}_\ep,
}
with the first order expansion term $\oset{1}{W_\ep}$ satisfying
\lalign{W1eqn}{
\oset{1}{b}_\ep{}^{0}\del_t \oset{1}{W}_\ep &= \frac{1}{\ep} c^I\del_I \oset{1}{W}_\ep
+ \oset{0}{b^I}\del_I\oset{1}{W}_\ep +
\oset{0}{b^I}\del_I\oset{1}{\omega} + \oset{1}{\Bt}(\ovec{0}{U},\ovec{1}{W},\ovec{0}{Y})
-\oset{1}{B}{}^0(\ovec{0}{W},\ovec{1}{X})+\oset{1}{\Fc}(\ovec{0}{U},\ovec{1}{W})
\label{W1eqn.1},\\
\oset{1}{W}_\ep(0) &= \oset{1}{\uset{0}{W}}-\oset{1}{\omega}(0), \label{W1eqn.2}
}
where
\gath{XYdefA}{
\oset{1}{b}_\ep{}^0 = b^0(0,\ep \oset{0}{W}) \, ,\\
\ovec{0}{U} = \oset{0}{U} \, , \quad
\ovec{0}{W} = \oset{0}{W}\, , \quad
\ovec{0}{X} = \del_t \oset{0}{W}\, , \quad
\ovec{0}{Y} = \del_I \oset{0}{W} \, ,
\intertext{and}
\ovec{1}{W} = (\oset{0}{W},\oset{1}{\omega}+\oset{1}{W_\ep})\, , \quad
\ovec{1}{X} = (\del_t \oset{0}{W},\del_t\oset{1}{\omega})\, .
}
Observe that
\eqn{b0}{
\oset{1}{b_\ep}{}^0 = \id,
}
by Proposition \ref{eplimD}. Substituting \eqref{Z2eqn} in \eqref{wsysdef} yields
\eqn{Z2eqnA}{
\oset{1}{\Bb}_\ep{}^0 +
\ep \oset{1}{b}_\ep{}^0\del_t\oset{1}{W}_\ep +
\ep^2 \frac{b^0_\ep-\oset{1}{b}_\ep{}^0}{\ep^2}\ep \del_t
\oset{1}{W}_\ep+
\ep^2 b^0_\ep\oset{2}{Z}_\ep = \frac{1}{\ep}c^I\del_I \oset{0}{W} + c^I\del_I \oset{1}{\omega}
+ c^I\del_I \oset{1}{W_\ep}
+ \ep c^I \del_I \oset{2}{Z_\ep} + \oset{1}{\Bb_\ep}
+ \ep^2 b^I_\ep \del_I\oset{2}{Z_\ep} + \Fc_\ep,
}
where
\lalign{Z2eqnB}{
b^0_\ep &= b^0(\ep^2 U,\ep W_\ep)\, , \label{Z2eqnB.1}\\
\oset{1}{\Bb}_\ep &= B\bigl(\ep,U,W_\ep,\del_I \oset{0}{W}+\ep
(\del_I\oset{1}{\omega}+\del_I\oset{1}{W_\ep}) \bigr)\, , \label{Z2eqnB.2}
\intertext{and}
\oset{1}{\Bb}_\ep{}^0 &= B^0(\ep^2U,\ep W_\ep,\del_t\oset{0}{W}+
\ep \del_t \oset{1}{\omega}) \label{Z2eqnB.3}\, .
}
Using \eqref{W1eqn.1}-\eqref{W1eqn.2}, we then find that
$\oset{2}{Z_\ep}$ satisfies
\lalign{Z2eqnC}{
b^0_\ep \del_t \oset{2}{Z_\ep} & = \frac{1}{\ep} c^I\del_I\oset{2}{Z_\ep} + b^I_\ep \del_I \oset{2}{Z_\ep}
+ \oset{2}{\Kcb_\ep}, \label{Z2eqnC.1}\\
\oset{2}{Z_\ep}(0) &=
\frac{W_\ep(0)-\oset{0}{W}(0)-\ep\bigl(\oset{1}{W_\ep}(0)+ \oset{1}{\omega}(0)\bigr)}{\ep^2}, \label{Z2eqnC.2}
}
where
\lalign{Z2eqnD}{
\oset{2}{\Kcb_\ep} = \frac{\oset{1}{b_\ep}{}^0- b^0_\ep }{\ep^2}& \ep \del_t\oset{1}{W_\ep} +
\frac{1}{\ep^2}\left[ \Bigl(\sum_{q=0}^{1}\ep^q \oset{q}{B}{}^0\bigl(\ovec{q-1}{W},
\ovec{q}{X} \bigr)- \oset{1}{\Bb}_\ep{}^0\Bigr) + \right.  \notag \\
& \left. \Bigl( \oset{1}{\Bb}{}_\ep - \sum_{q=0}^{1}\ep^q \oset{q}{B}\bigl(\ovec{q-1}{U},\ovec{q}{W},
\ovec{q}{Y} \bigr)\Bigr)  + \Bigl( \Fc_\ep - \sum_{q=0}^{1}\ep^q \oset{q}{\Fc}\bigl(\ovec{q-1}{U},\ovec{q}{W} \bigr)
\Bigr)
\right], \label{Z2eqnD.1}
}
and
\eqn{XYdefB}{
\ovec{1}{U}  = (\oset{0}{U},\oset{1}{U})\, ,\quad
\ovec{1}{Y} = (\del_I \oset{0}{W},\del_I \oset{1}{\omega}+\del_I\oset{1}{W}_\ep) \, . }
Letting
\eqn{Z2eqnEa}{
\ovec{1}{\tilde{X}} = (\del_t\oset{0}{W},\del_t\oset{1}{\omega}+\del_t\oset{1}{W}_\ep),
}
it follows from Proposition \ref{eplimC} that
\leqn{Z2eqnE}{
\oset{2}{\Kcb}_\ep = \Lc\bigl(\ovec{1}{U},\ovec{1}{W},\ovec{1}{\tilde{X}},\ovec{1}{Y},\oset{2}{Z_\ep}
\bigr) +
\Mc\bigl(\ep,U,\ovec{1}{W},\ovec{1}{\tilde{X}},\ovec{1}{Y},\ep\oset{2}{Z_\ep} \bigr)
}
for analytic maps $\Lc$ and $\Mc$ with  $\Lc$ linear in $\oset{2}{Z_\ep}$.

As we shall see in Theorem \ref{foeA}, when the initial data is chosen
such that $\norm{\del^2_tW_\ep(0)}_{H^{k-2}_\delta}$ remains
bounded as $\ep \searrow 0$,  the $\ep$ dependence can be removed
from the first order expansion coefficient. This is accomplished by
replacing \eqref{W1eqn.1}-\eqref{W1eqn.2} with  a related, but different
$\ep$ independent
version. To describe this system, we  let
\eqn{W1def}{
\oset{1}{W}=(\oset{1}{W}_4{}^{ij},\oset{1}{W}{}^{ij}_I,
\delta\oset{1}{\uf}{}^{ij},\oset{1}{\alpha},\oset{1}{w}{}^i)^T\, ,
}
and define
projection operators by
\eqn{Pi4def}{
\Pi_4(\oset{1}{W}) = (\oset{1}{\uf}{}^{ij}_4)\,
\AND
\Pi_J(\oset{1}{W})=  (\oset{1}{W}{}^{ij}_J) \, .
}
Then the system that replaces \eqref{W1eqn.1}-\eqref{W1eqn.2} is:
\lalign{W1eqnA}{
\del_t \oset{1}{W} &=
\oset{0}{b^I}\del_I\oset{1}{W_\ep} +
\oset{0}{b^I}\del_I\oset{1}{\omega} + \oset{1}{\Bt}(\ovec{0}{U},\ovec{1}{W},\ovec{0}{Y})
-\oset{1}{B}{}^0(\ovec{0}{W},\ovec{1}{X})+
\oset{1}{\Fc}(\ovec{0}{U},\ovec{1}{W})+ c^I\del_I\oset{2}{\omega}
\label{W1eqnA.1} \, ,\\
\oset{1}{W}(0) &= \oset{1}{\uset{0}{W}}-\oset{1}{\omega}(0) \, ,
\label{W1eqnA.2}
}
where
\lalign{W1eqnA2}{
\oset{2}{\omega} & = (\oset{2}{\omega}_4{}^{ij},\del_I \oset{2}{\Omega}{}^{ij},
0,0,0)^T \, , \label{W1eqnA2.1} \\
\Delta \oset{2}{\omega}_4 &= -\del^J\Pi_J\bigl(\oset{1}{\Bt}(\ovec{0}{U},\ovec{1}{W},\ovec{0}{Y})
-\oset{1}{B}{}^0(\ovec{0}{W},\ovec{1}{X})+
\oset{1}{\Fc}(\ovec{0}{U},\ovec{1}{W})
\bigr)\, , \label{W1eqnA2.2}
\intertext{and}
\Delta \oset{2}{\Omega} &= - \Pi_4\bigl(
\oset{1}{\Bt}(\ovec{0}{U},\ovec{1}{W},\ovec{0}{Y})
-\oset{1}{B}{}^0(\ovec{0}{W},\ovec{1}{X})+
\oset{1}{\Fc}(\ovec{0}{U},\ovec{1}{W})
\bigr)\, . \label{W1eqnA2.3}
}
Existence of solutions to
the initial value problem \eqref{W1eqnA.1}-\eqref{W1eqnA.2} is
covered by the following Proposition.
\begin{prop} \label{foeB} \mnote{[foeB]}
Let $\delta$, $k$, $s$, $K_1$, $R$, $\Rb$, and $\tau$ be as in Proposition \ref{eplimA},
$T_0^M$ be as in Proposition \ref{cogA}, and suppose $T_0 < T^M_0$.
If $s$ and $\tau$ are chosen so that
$s\geq 1$, and $16\tau$ $>$ $\max\{32K_1,T_0\sup_{0\leq t \leq T_0}\sup\norm{\wh^I(t)}_{L^\infty}\}$,
then there exists a map
\eqn{foeB1}{
\oset{1}{W} \in X_{T_0,s-1,k-1,\delta}
}
such that $\oset{1}{W}(t)$ is the unique solution
to the initial value problem \eqref{W1eqnA.1}-\eqref{W1eqnA.2}, and
\eqn{foeB1a}{
\supp\, \oset{1}{\rho}(t) \subset B_{\bar{R}} \quad
\text{for $0\leq t < T_0$,}
}
where $\oset{1}{\rho}=
\frac{2n}{(4Kn(n+1))^n} \alphah^{2n-1} \oset{1}{\alpha}$.
Moreover, if the initial data satisfies
$c^I\del_I \oset{1}{W}(0)= 0$,
then
\eqn{foeB2}{
c^I\del_I \oset{1}{W}(t)=0 \quad \text{for $0\leq t < T_0$},
\AND
\oset{2}{\omega}_I, \; \oset{2}{\omega}_4 \in X_{T_0,s-1,k-1,\delta-1}.
}
\end{prop}
\begin{proof}
By construction, we have
\leqn{foeB5}{
\oset{1}{W_0}-\oset{1}{\omega}(0)\in H^{k-1}_{\delta-1}\, .
}
Next, we observe that the map
\lalign{foeB6a}{
H^{\ell+1}_\delta \times \bigl(H^{\ell}_{\delta-1}\times
& H^{\ell-1}_{\delta-1}\bigr)\times H^{\ell-1}_{\delta-1}
\times H^{\ell-1}_{\delta-1} \ni (\ovec{0}{U},\ovec{1}{W},\ovec{0}{X},\ovec{0}{Y})
\notag \\
& \longmapsto \Pi_4\bigl(
\oset{1}{\Bt}(\ovec{0}{U},\ovec{1}{W},\ovec{0}{Y})
-\oset{1}{B}{}^0(\ovec{0}{W},\ovec{1}{X})+
\oset{1}{\Fc}(\ovec{0}{U},\ovec{1}{W})
\bigr) \in H^{\ell-1}_{\delta-2} \label{foeB6}
}
is analytic for $\ell > 3/2+1$, which follows directly from the
weighted estimates of Appendix \ref{winq} (see also
Appendix A of \cite{Oli06}).
It therefore follows that the
system \eqref{W1eqnA.1}-\eqref{W1eqnA2.3}
satisfies all the hypotheses of Theorem
\ref{hypA}. Thus,  there exists a unique solution
\leqn{foeB7}{
\oset{1}{W}\in X_{T_0,s-1,k-1,\delta-1}
}
satisfying the initial value problem \eqref{W1eqnA.1}-\eqref{W1eqnA.2}.
Furthermore, from \eqref{foeB4}-\eqref{foeB6}, it is clear that $\oset{2}{\omega}_I =
\del_I\oset{2}{\Omega} \in X_{T_0,s-1,k,\delta-1}$.
Note that we have used the linearity of the system
\eqref{W1eqnA.1}-\eqref{W1eqnA2.3} in $\oset{1}{W}$ to
conclude that the solution can be continued as long
as the coefficients are well defined, which is the
case for $0\leq t \leq T_0< T^M_0$.

By assumption, the initial data satisfies
\leqn{foeB10}{
c^I \del_I \oset{1}{W}(0) = 0,
}
while from Proposition \ref{eplimD}
we have that
\leqn{foeB16}{
\oset{0}{W}_4{}^{ij}(t)= \oset{0}{W}_I{}^{ij}(t)=
\delta\oset{0}{\uf}{}^{ij}(t) = 0,
}
and hence
\leqn{foeB17}{\quad
\oset{0}{\uf}_4{}^{ij}(t) = 0,
\quad
\oset{0}{\uf}_I{}^{ij}(t) = \delta^i_4\delta^j_4
\del_I\Phih(t),
\AND \oset{0}{\uf}{}^{ij}(t)  = 0.
}
From this it follows that $\oset{0}{b}{}^I$
has a block diagonal structure of the form
\eqn{foeB11}{
\oset{0}{b}{}^I = \begin{pmatrix} 0 & 0 \\
0 & *
\end{pmatrix},
}
and consequently
\leqn{foeB12}{
\oset{0}{b}{}^I \del_I \oset{1}{\omega}= 0\, , \quad
\Pi_4\bigl(\oset{0}{b}{}^I \del_I \oset{1}{W}\bigr)= 0\, , \AND
\Pi_J\bigl(\oset{0}{b}{}^I \del_I \oset{1}{W}\bigr)= 0 \, .
}
Next, a straight forward calculation using \eqref{W1eqnA.1},
\eqref{W1eqnA2.1}-\eqref{W1eqnA2.2}, and
\eqref{foeB12} shows that
$\del_t\bigl(c^I\del_I\oset{1}{W}\bigr) = 0$, and hence
\eqn{foeB14}{
c^I\del_I\oset{1}{W}(t) = 0
\quad \text{for $0\leq t < T_0$}
}
by \eqref{foeB10}. By the definition of the $c^I$, this is equivalent
to (since $\delta <0$)
\leqn{foeB15}{
\oset{1}{W}_4(t) = 0 \AND \del^I\oset{1}{W}_I(t) = 0.
}
A short calculation using \eqref{W1eqnA.1} and \eqref{foeB15} then shows
that
\leqn{foeB18}{
\del_t \delta\oset{1}{\uf}{}^{ij} = \oset{1}{\omega}_4{}^{ij}=
\delta^i_4\delta^j_4\del_t\Phih\, .
}
However, $\delta\oset{1}{\uf}{}^{ij}(0)=0$ (see Proposition \ref{idat}),
and so integrating
\eqref{foeB18} yields
\leqn{foeB19}{
\delta\oset{1}{\uf}{}^{ij}= \delta^{i}_4\delta^j_4\bigl(\Phih(t)-\Phih(0)
\bigr)\, ,
}
and
\leqn{foeB20}{
\oset{1}{\uf}{}^{ij}(t) = \oset{0}{\uset{0}{\ufb}}{}^{ij}+\delta\oset{1}{\uf}{}^{ij}(t)
= \delta^i_4\delta^j_4 \Phih(t) \, .
}
Also by \eqref{foeB15}, we have that
\leqn{foeB21}{
\oset{1}{\uf}_4{}^{ij}(t) =\oset{1}{\omega}_4{}^{ij}(t)
+ \oset{1}{W}_4{}^{ij}(t)= \delta^i_4\delta^j_4
\Phih(t),
}
while
\leqn{foeB22}{
\oset{1}{\uf}_I{}^{ij}(t) = \oset{1}{\omega}_I{}^{ij}(t)
+ \delta^i_4\delta^j_4\del_I\oset{1}{\Phi}(t)+ \oset{1}{W}_I{}^{ij}(t),
}
where
\leqn{foeB23}{
\Delta \oset{1}{\Phi} = \oset{1}{\rho}\, .
}
We remark that in obtaining
\eqref{foeB23}, we have used
$\text{supp}\, \oset{1}{\rho}(t) \subset B_{\Rb}$ for $0\leq t < T_0$,
which follows from the definition of $\oset{1}{\rho}$
and Proposition \ref{cogA}.

Using \eqref{foeB16}, \eqref{foeB17}, \eqref{foeB15}, \eqref{foeB20},
\eqref{foeB21}, and \eqref{foeB22} together,
we can write \eqref{W1eqnA2.2} as
\leqn{foeB26}{
\Delta \oset{2}{\omega}_4{}^{ij}= \del^J\left(
\del_t \oset{1}{\omega}_J{}^{ij}+ \delta^i_4\delta^j_4\del_J\del_t\oset{1}{\Phi}
\right)\, .
}
Moreover, it follows from the evolution equation \eqref{W1eqnA.1} that
\leqn{foeB27}{
\del_t\oset{1}{W}_I{}^{ij}= -\del_t\oset{1}{\omega}_I{}^{ij}
+ \del_I \oset{2}{\omega}_4{}^{ij}- \delta^{i}_4\delta^{j}_4\del_I\del_t\oset{1}{\Phi}\, .
}
We also note that
\leqn{foeB28}{
\oset{1}{\omega}_I{}^{ij} = \del_I\Delta^{-1}\bigl(2\rhoh\delta^{(i}_J\delta^{j)_4}\wh^J
\bigr),
}
by Proposition \eqref{eplimD}, and hence
\leqn{foeB29}{
\del_t\del_{[J}\oset{1}{W}_{I]}{}^{ij} = 0
}
by \eqref{foeB27}.
However, $\del_{[J} \oset{1}{W}_{I]}(0)=0$ by Proposition \ref{idat}, and
thus we
get from \eqref{foeB29} that
$\del_{[J}\oset{1}{W}_{I]}{}^{ij}(t) = 0$.
This combined with \eqref{foeB15} shows that (since $\delta < 0$)
\leqn{foeB31}{
\oset{1}{W}_I{}^{ij}(t) = 0 \, ,
}
and hence
\leqn{foeB32}{
\oset{1}{\uf}_I{}^{ij}(t) = \oset{1}{\omega}_I{}^{ij}(t)+
\delta^i_4\delta^j_4\del_I\oset{1}{\Phi}(t)\, .
}
Using \eqref{foeB16}, \eqref{foeB17}, \eqref{foeB21}, \eqref{foeB31}, \eqref{foeB32},
and the evolution equation \eqref{W1eqnA.1}, a straightforward calculation
then shows that the pair $\{\oset{1}{\alpha},\oset{1}{w}{}^I\}$ satisfy
\lalign{foeB33}{
\del_t\oset{1}{\alpha} - \wh^I\del_I\oset{1}{\alpha} -\frac{\alphah}{2n}
\del_I\oset{1}{w}{}^I -\oset{1}{w}{}^I\del_I\alphah -\frac{\oset{1}{\alpha}}{2n}
\del_I\wh^I & = 0\, , \label{foeB33.1}\\
\del_t\oset{1}{w}{}^J-\wh^I\del_I \oset{1}{w}{}^J - \frac{\alphah}{2n}\del^J\oset{1}{\alpha}
-\del^J\oset{1}{\Phi}
- \oset{1}{w}^I\del_I\wh^J - \frac{\oset{1}{\alpha}}{2n}\del^J\alphah &= 0\, .
\label{foeB33.2}
}
Also, we observe that
\leqn{foeB34}{
\oset{2}{\omega}_4 = 2\delta^{(i}_J\del^{j)}_4\del_t\bigl(\rhoh\wh^J\bigr)
+ \delta^i_4\delta^j_4\del_t\oset{1}{\Phi},
}
by \eqref{foeB26} and \eqref{foeB28}, and that
\lgath{foeB35}{
\del_I \bigl(\oset{1}{w}{}^I\rhoh\bigr) = \frac{2n\alphah^{2n-1}}{(4Kn(n+1))^n}
\left[ \oset{1}{w}{}^I\del_I\alphah + \frac{\alphah}{2n} \del_I \oset{1}{w}{}^I
\right], \label{foeB35.1}
\intertext{and}
\del_I\bigl(\wh^I\oset{1}{\rho}\bigr) = \frac{2n\alphah^{2n-1}}{(4Kn(n+1))^n}
\left[
\oset{1}{\alpha}\del_I\wh^I+ \wh^I\del_I\oset{1}{\alpha}
\right]
+ \frac{2n(2n-1)}{(4Kn(n+1))^n}\bigl(\wh^I\del_I\alphah\bigr)\alphah^{2n-2}
\oset{1}{\alpha} \, . \label{foeB35.2}
}
But, by \eqref{newtA.1}, we have
\leqn{foeB36}{
\del_t\oset{1}{\rho} = \frac{2n}{(4Kn(n+1))^n}\left[
\del_t\oset{1}{\alpha} + \frac{2n-1}{2n}\oset{1}\del_I\wh^I
\right] + \frac{2n(2n-1)}{(4Kn(n+1))^n}\bigl(\wh^I\del_I\alphah\bigr)\alphah^{2n-1}\oset{1}{\alpha},
}
and
therefore
\eqn{foeB37}{
\del_t\oset{1}{\rho} - \del_I\bigl(\oset{1}{w}{}^I\rhoh+ \wh^I\oset{1}{\rho}\bigr) = 0
}
by \eqref{foeB33.1}, \eqref{foeB35.1}, \eqref{foeB35.2}, and \eqref{foeB36}. It then follows from
\eqref{foeB23} that
\eqn{foeB38}{
\del_t \oset{1}{\Phi} = \del_I \Delta^{-1}\bigl( \oset{1}{w}{}^I\rhoh + \wh^I\oset{1}{\rho}\bigr),
}
and hence
\eqn{foeB39}{
\oset{2}{\omega}_4{}^{ij} \in X_{T_0,s-1,k,\delta-1}
}
by \eqref{foeB7}, \eqref{foeB34}, and Proposition \ref{cogA}.
\end{proof}

\begin{thm} \label{foeA} \mnote{[foeA]}
Let $\delta$, $k$, $s$, $K_1$, $R$, $\Rb$, $\tau$, $T$, and $W_\ep(t)$ be as in
Proposition \ref{eplimA},
$\{\oset{0}{W}(t),\oset{1}{\omega}(t)\}$ as in Proposition \ref{eplimD},
$T_0^M$ as in Proposition \ref{cogA}, and suppose $T_0 < T^M_0$.
If $s$ and $\tau$ are chosen so that
$s\geq 2$, and $16\tau$ $>$ $\max\{32K_1,T_0\sup_{0\leq t \leq T_0}\sup\norm{\wh^I(t)}_{L^\infty}\}$,
then for $\ep_0>0$ small enough,
\begin{itemize}
\item[(i)] there exist constants $K_2$, $K_3$ such that the
solution $W_\ep(t)$ $(0<\ep \leq \ep_0)$ exists on the interval
$[0,\tilde{T}_\ep)$, where
\eqn{foeA1.1}{
\tilde{T}_\ep = \min\left\{T_0,\frac{1}{K_2}\ln\left(\frac{K_3}{\ep}\right)
\right\},
}
and obeys the bounds
\gath{foeA1.2}{
\sup_{0\leq t < \tilde{T}_\ep}\max\{\norm{\ep\ufb_\ep(t)}_{L^\infty},\norm{\ep\alpha_\ep(t)},
\norm{\ep w^i_\ep(t)}_{L^\infty}\} < 2K_0\, , \\
\sup_{0\leq t < \tilde{T}_\ep}\norm{W_\ep(t)}_{W^{1,\infty}} < \infty,
\quad
\supp\, \rho_\ep(t) \subset B_{\Rb}\, ,
}
\item[(ii)] and there exists maps
\eqn{foeA1}{
\oset{1}{W_\ep} \in  X_{T_0,s-1,k-1,\delta-1} \quad  0<\ep\leq \ep_0,
}
such that $\oset{1}{W_\ep}$ is the unique solution to the initial value problem \eqref{W1eqn.1}-\eqref{W1eqn.2},
and
\eqn{foeA2a}{
\norm{W_\ep(t)-\oset{0}{W}(t)-\ep \bigl(\oset{1}{\omega}(t)+\oset{1}{W}_\ep(t)\bigr)}_{H^{k-2}}
\lesssim
\norm{W_\ep(t)-\oset{0}{W}(t)-\ep \bigl(\oset{1}{\omega}(t)+\oset{1}{W}_\ep(t)\bigr)}_{H^{k-2}_{\delta-1,\ep}}
\lesssim e^{K_2t}\ep^2\,
}
for all $(t,\ep) \in [0,\tilde{T}_\ep)\times (0,\ep_0]$.
\item[(iii)]
Moreover, if $W_\ep(0)$ satisfies
$\norm{\del_t^2 W_\ep(0)}_{H^{k-2}_{\delta-1}}$ $\lesssim 1$ for  $0\leq \ep \leq \ep_0$,
then
\eqn{foeA2}{
\norm{W_\ep(t)-\oset{0}{W}(t)-\ep \bigl(\oset{1}{\omega}(t)+\oset{1}{W}(t)\bigr)}_{H^{k-2}}
\lesssim
\norm{W_\ep(t)-\oset{0}{W}(t)-\ep \bigl(\oset{1}{\omega}(t)+\oset{1}{W}(t)\bigr)}_{H^{k-2}_{\delta-1,\ep}}
\lesssim e^{K_2t}\ep^2
}
for all $(t,\ep) \in [0,\tilde{T}_\ep)\times (0,\ep_0]$,
where $\oset{1}{W}\in X_{T_0,s-1,k-1,\delta-1}$ is the unique solution to the
initial value problem \eqref{W1eqnA.1}-\eqref{W1eqnA.2}.
\end{itemize}
\end{thm}
\begin{proof}
\textbf{(i)-(ii):} Fix $T_* < \min\{T,T_0\}$, and let
\alin{foeA7}{
C_1 & = \sup_{0\leq t \leq T_*}\norm{\oset{0}{W}(t)}_{H^k_{\delta-1}}
+ \sup_{0\leq t \leq T_*}\norm{\del_t \oset{0}{W}(t)}_{H^{k-1}_{\delta-1}}, \\
C_2 &= \sup_{0\leq t \leq T_*}\norm{\oset{1}{\omega}(t)}_{H^k_{\delta-1}}
 + \sup_{0\leq t \leq T_*}\norm{\del_t\oset{1}{\omega}(t)}_{H^{k-1}_{\delta-1}},
\intertext{and}
C_3 & = \norm{\oset{1}{\uset{0}{W}}-\oset{1}{\omega}(0)}_{H^{k-1}_{\delta-1}}.
}
Since
\eqn{foeA8}{
\norm{\ufbo^{ij}_\ep}_{H^{k+1}_{\delta}}
\leq \frac{K_0}{\sqrt{\ep_0}\Csob}\, ,
}
and  $\oset{1}{W_\ep}$ satisfies the linear equation
\eqref{W1eqn.1}, it follows from
the energy estimates derived in the proof of Theorem \ref{hypA} that
there exists a constant $K_2=K_2(C_1,C_2,K_0/(\sqrt{\ep_0}\Csob))$ such that
\leqn{foeA9}{
\norm{\oset{1}{W_\ep}(t)}_{H^{k-1}_{\delta-1,\ep}}
\leq e^{K_2T_*}C_3 + K_2 \quad \forall \; (t,\ep)\in
[0,T_*]\times (0,\ep_0]\, .
}
Next, we observe that
\lalign{foeA10}{
\norm{\ep\ufb^{ij}_\ep(t)}_{L^\infty}
&\leq \Csob\left[ \ep\norm{\ufbo^{ij}_\ep}_{H^{k+1}_{\delta}} +
\norm{W_\ep(t)-\oset{0}{W}(t)}_{H^{k-2}_{\delta-1,\ep}}\right] &&\text{(by \eqref{sobolev})}
\notag \\
& \leq K_0 + \ep\Csob\bigl[\ep\norm{\oset{2}{Z}_\ep(t)}_{H^{k-2}_{\delta-1,\ep}}
+ \norm{\oset{1}{W}_\ep(t)}_{H^{k-1}_{\delta-1,\ep}}+ C_2 \bigr] \, , \label{foeA10.1}\\
\norm{W_\ep(t)}_{W^{1,\infty}}
& \leq \Csob\bigl[\ep^2\norm{\oset{2}{Z}_\ep(t)}_{H^{k-2}_{\delta-1,\ep}}
+ \ep\norm{\oset{1}{W}_\ep(t)}_{H^{k-1}_{\delta-1,\ep}}+\ep C_2+C_1 \bigr],
\label{foeA10.2}
\intertext{and}
\norm{W_\ep(t)-\oset{0}{W}_\ep(t)}_{W^{1,\infty}}
& \leq \Csob\bigl[\ep^2\norm{\oset{2}{Z}_\ep(t)}_{H^{k-2}_{\delta-1,\ep}}
+ \ep\norm{\oset{1}{W}_\ep(t)}_{H^{k-1}_{\delta-1,\ep}}+\ep C_2\bigr].
\label{foeA10.3} }
Setting $\oset{2}{\Zc}_{\ep}(t) = \ep \oset{2}{Z}_{\ep}(t)$, we
note that by construction there exists a constant
$C_5$ such that
$\norm{\oset{2}{\Zc}_{\ep}(0)}_{H^{k-2}_{\delta-1,\ep}}$ $\leq$ $\ep C_4$. 
Moreover, from the error equation \eqref{Z2eqnC.1},
it is clear that $\oset{2}{\Zc}$ satisfies an equation
to which Theorem \ref{hypA} applies. Therefore, for
any $K_3 > \ep C_4$ $(0\leq \ep \leq \ep_0)$ there
exists constants $K_4,K_5$ such that $\oset{2}{\Zc}(t)$
satisfies an estimate of the form
\leqn{foeA13}{
\norm{\oset{2}{\Zc}(t)} \leq \ep \left(e^{K_4t}[C_4+K_5]-K_5
\right) \leq K_3 \quad \text{for $0\leq t < \tilde{T}$,}
}
where
\leqn{foeA14}{
\tilde{T} = \min\left\{T_*,\frac{1}{K_4}
\ln\left(\frac{K_3+\ep K_5}{\ep(C_4+K_5)}\right) \right\}\, .
}
Statements (i) and (ii) now follow directly from
Propositions \ref{eplimA} and \ref{cogA}, and the
estimates \eqref{imbedB}, \eqref{foeA10.1}-\eqref{foeA10.3}, \eqref{foeA13},
and \eqref{foeA14}, provided $\ep_0$ is chosen small enough.
\bigskip

\noindent \textbf{(iii):} To prove statement (iii), we first observe that it
follows from the evolution equation \eqref{wsysdef} that
the condition
$\norm{\del_t^2W_\ep(0)}_{H^{k-2}_{\delta-1}}\lesssim 1$ for
$0 <\ep \leq \ep_0$ is equivalent to the
condition $c^I\del_I\oset{1}{W}(0)=0$. Then
replacing $\oset{1}{W}_\ep(t)$, and
$\oset{2}{Z}_\ep(t)$ in \eqref{Z2eqn} with
$\oset{1}{W}(t)$, and $\oset{2}{\omega}+\oset{2}{Z}_\ep(t)$,
respectively, it is not difficult using Proposition \ref{foeB}
to show that the new error term $\oset{2}{Z}_\ep(t)$ will satisfy
the same type of estimate as above. We emphasize
that the key property used to make  this replacement is
that $\oset{1}{W}(t)$ and $\oset{2}{\omega}(t)$ satisfy
$c^I\del_I\oset{1}{W}(t)=0$ and
$\oset{2}{\omega} \in X_{T_0,s-1,k-1,\delta-1}$.
The proof of statement (iii) now follows as we
are able to
replace $\oset{1}{W}_\ep(t)$ with $\oset{1}{W}(t)$ 
everywhere in the above estimates.
\end{proof}

\sect{hoe}{Higher order expansions and convergence}

\begin{thm} \label{hoeA} \mnote{[hoeA]}
Let $\delta$, $k$, $s$, $K_1$, $R$, $\Rb$, and $W_\ep(t)$ be as in 
Proposition \ref{eplimA},
$\{\oset{0}{W}(t),\oset{1}{\omega}(t)\}$ as in Proposition \ref{eplimD},
$T_0^M$ as in Proposition \ref{cogA}, $\oset{1}{W}_\ep(t)$ and $\tau$ as in 
Theorem \ref{foeA}, and suppose $T_0 < T^M_0$.
If $s\geq 3$, then for $\ep_0$ small enough, there exists an
infinite sequence of maps
\eqn{hoeA1}{
\oset{q}{W}_\ep \in X_{T_0,s-2,k-2,\delta-1}   \quad q\in \Zbb_{\geq 2}
}
such that
\begin{itemize}
\item[(i)] each $\oset{q}{W}(t)$ satisfies a linear (non-local) symmetric
hyperbolic system with
initial data $\oset{q}{W}_\ep(0)= \oset{q}{\uset{0}{W}}$ and coefficients depending on $\ep$, $\oset{0}{W}$, $\oset{1}{\omega}$,
$\oset{r}{U}$ for $0\leq r\leq q$, and $\oset{r}{W}_\ep$ for $1\leq r\leq q-1$,
\item[(ii)]
\eqn{hoeA2}{
\norm{\oset{q}{W}_\ep(t)}_{H^{k-2}} +
\ep\norm{\del_t\oset{q}{W}_\ep(t)}_{H^{k-3}} \lesssim
\norm{\oset{q}{W}_\ep(t)}_{H^{k-2}_{\delta-1,\ep}} +
\ep\norm{\oset{q}{\del_t  W_\ep}}_{H^{k-3}_{\delta-1,\ep}} \lesssim
1 }
for all $(t,\ep,q)\in [0,T_0)\times (0,\ep_0]\times \Zbb_{\geq 2}$, and
\item[(iii)] 
\leqn{hoeA3}{
W_\ep(t) = \oset{0}{W}(t) + \ep(\oset{1}{\omega}(t)+\oset{1}{W_\ep})
+ \sum_{q=0}^\infty \ep^q \oset{q}{W_\ep}(t) \quad (t,\ep)
\in [0,T_0)\times (0,\ep_0],
}
where the sum converges uniformly in $C^0([0,T_0);H^{k-3}_{\delta-1,\ep})$ and
$C^0([0,T_0);H^{k-3})$.  
\item[(iv)]
Moreover, if $s-2\geq p \geq 1$, and the initial
data is chosen so that
\eqn{hoeA4}{
\norm{\del_t^{q+1}W_\ep(0)}_{H^{k-(q+1)}_{\delta-1}} \lesssim 1  \quad q=1,2,\ldots,p,
}
then there exists $\ep$-independent maps
\eqn{hoeA5}{
\oset{q}{W} \in X_{T_0,s-q.k-q,\delta-1}
 \AND \oset{q+1}{\omega} \in X_{T_0,s-q,k-q,\delta-1}\quad q = 1,2,\ldots,p
}
such that
\begin{itemize}
\item[(iv.a)] each $\oset{q}{W}$ satisfies a $\ep$-independent 
linear (non-local) symmetric
hyperbolic  system with coefficients depending only on 
$\oset{r}{U}$ for $0\leq r\leq q$, $\oset{r}{\omega}$ for $0\leq r \leq q+1$, and 
$\oset{r}{W}$ for $0\leq r\leq q-1$, and
\item[(iv.b)] the terms  $\oset{q}{W}_\ep$ in the sum
\eqref{hoeA3} can be replaced by $\oset{q}{\omega}+\oset{q}{W}$ for $1\leq q \leq p$
with the sum converging uniformly 
 $C^0([0,T_0),H^{k-(q+2)}_{\delta-1,\ep})$ and
$C^0([0,T_0);H^{k-(q+2)})$.
\end{itemize}
\end{itemize}
\end{thm}
\begin{proof}
The proof of this Theorem follows from a straightforward adaptation of the proof of Theorem 3 in \cite{Scho88}.
We will only sketch the details. 

Following Schochet \cite{Scho88} (see also \cite{KM82}), we consider the following  iteration:
\lalign{ZmeqnA}{
b^0(\oset{m}{Z_\ep}) \del_t \oset{m+1}{Z_\ep} & = \frac{1}{\ep} c^I\del_I\oset{m+1}{Z_\ep} + b^I(
 \oset{m}{Z_\ep})\del_I \oset{m+1}{Z_\ep}
+\Lc(\oset{m+1}{Z_\ep}) + \ep \Mc(\oset{m}{Z_\ep}), \label{ZmeqnA.1}\\
\oset{m+1}{Z_\ep}(0) &= \sum_{q=2}^{m+1}\ep^{q-2}\oset{q}{\uset{0}{W}}, \label{ZmeqnA.2}
}
where
\gath{ZmeqnB}{
Z_1 = 0,\quad 
\oset{m}{\Wb_\ep} = \oset{0}{W}+\ep(\oset{1}{\omega}+\oset{1}{W_\ep}) +\ep^2 \oset{m}{Z_\ep},\quad
b^I(\oset{m}{Z_\ep}) = b^I(\oset{m}{\Wb_\ep},\ep U,\ep \oset{m}{\Wb_\ep},\ep^2 U), \\
b^0(\oset{m}{Z_\ep}) = b^0(\ep^2 U, \ep \oset{m}{\Wb_\ep}),\quad \Lc(\oset{m+1}{Z_\ep}) = \Lc\bigl(\ovec{2}{U},\ovec{1}{W},\ovec{1}{\tilde{X}},
\ovec{1}{Y},\oset{m+1}{Z_\ep}\bigr), \AND
\Mc(\oset{m}{Z_\ep}) = \Mc\bigl(\ep,U,\ovec{1}{W},\ovec{1}{\tilde{X}},\ovec{1}{Y},\oset{m}{Z_\ep} \bigr).
}
Using the energy estimates of Theorem \ref{hypA} and the weighted Sobolev estimates in Appendix \ref{winq}  (see also \cite{Oli06}), it is clear
the arguments of Schochet can be generalized to
show that
\leqn{Zmest}{
\norm{\oset{m}{Z}(t)}_{H^{k-2}_{\delta-1,\ep}}  + \ep \norm{\del_t\oset{m}{Z}(t)}_{H^{k-3}_{\delta-1,\ep}} \lesssim 1,
}
and
\leqn{Zmdiffest}{
\norm{\oset{m+2}{Z}(t)-\oset{m+1}{Z}(t)}_{H^{k-3}_{\delta-1,\ep}} \lesssim
\ep \norm{\oset{m+1}{Z}(t)-\oset{m}{Z}(t)}_{H^{k-3}_{\delta-1,\ep}} +
\ep^{m}\norm{\oset{m+2}{\uset{0}{W}}}_{H^{k-3}_{\delta-1,\ep}} \, 
}
for all $(t,\ep) \in [0,T_0)\times (0,\ep_0]$.
Therefore by \eqref{Wexp}, \eqref{Zmest}, \eqref{Zmdiffest}, and the uniqueness of solutions
to the evolution equation \eqref{ZmeqnA.1}, we see that for $\ep_0$ small
enough the sequence 
$\oset{0}{W}(t) + \ep\bigl(\oset{1}{W}_\ep(t) + \oset{1}{\omega(t)}\bigr) + \ep^2 \oset{m}{Z}_\ep(t)$
converges in $C^0([0,T_0),H^{k-3}_{\delta-1,\ep})$ to $W_\ep(t)$ for each $\ep\in (0,\ep_0]$.
Therefore, defining 
\eqn{Wmdef}{
\oset{m+1}{W}_{\ep}(t)  = \frac{ \oset{m+1}{Z}_{\ep}(t)-\oset{m}{Z}_{\ep}(t)}{\ep^{m-1}},
} we have that
\eqn{Wsum}{
W_\ep(t) = \oset{0}{W}(t) + \ep\bigl(\oset{1}{W}_\ep(t)+ \oset{1}{\omega(t)}\bigr) + \sum_{q=2}^\infty \ep^q \oset{q}{W}_\ep(t)
}
with the sum converging in $C^0([0,T_0),H^{k-3}_{\delta-1,\ep})$  for each $(\ep \in (0,\ep_0]$. Moreover,
because of the inequality \eqref{imbedB}, it follows that the sum converges uniformly in
$C^0([0,T_0),H^{k-3})$ for $\ep \in  (0,\ep_0]$. This completes the proof of statements (i)-(iii). The proof of
statement (iv) also follows easily from the arguments used in the proof of Theorem 3 in \cite{Scho88}.
\end{proof}

\begin{rem}\mnote{[pneqns]} \label{pneqns}
{\rm The equations satisfied by the $\oset{q}{W}$ from   
part (iv) of Theorem \eqref{hoeA} are:
\alin{pneqnsA}{
\del_t \oset{q}{W} &= \oset{0}{b}{}^I(\oset{0}{W})\del_I\oset{q}{W}
+   \oset{0}{b}{}^I(\oset{0}{W})\del_I\oset{q}{\omega}
+ \oset{q}{\tilde{B}}\bigl(\ovec{q-1}{U},\ovec{q}{W},
\ovec{q-1}{Y}\bigr)
- \del_t\oset{q}{\omega}\notag  \\
& \hspace{1.0cm} - \oset{q}{\tilde{B}}{}^0
\bigl(\ovec{q-2}{U},\ovec{q-1}{W},
\ovec{q-1}{X}\bigr)
+ \oset{q}{\Fc}\bigl(\ovec{q-1}{U},\ovec{q}{W}\bigr)
+ c^I\del_I \oset{q+1}{\omega}\, , \\
c^I\del_I\oset{q}{W} &= 0 \, , \\
\oset{q}{W}(0) &= \oset{q}{\uset{0}{W}}-\oset{q}{\omega}(0)\, ,
}
where
\gath{pneqnsB}{
\ovec{q}{U} = (\oset{0}{U},\ldots,\oset{q}{U} ,
\quad \ovec{q}{W}= (\oset{0}{W},\oset{1}{\omega}+\oset{1}{W},
\ldots,\oset{q}{\omega}+\oset{q}{W}), \\
\ovec{q}{X} = (\del_t\oset{0}{W},\del_t\oset{1}{\omega}+\del_t\oset{1}{W},
\ldots,\del_t\oset{q}{\omega}+\del_t\oset{q}{W}), 
\AND
\ovec{q}{Y} = (\del_I\oset{0}{W},\del_I\oset{1}{\omega}+\del_I\oset{1}{W},
\ldots,\del_I\oset{q}{\omega}+\del_I\oset{q}{W}) .
}
}
\end{rem}

\sect{fpne}{The first post-Newtonian expansion}

We are now ready to prove the main theorem that
guarantees the existence of a large class of
solutions to the Einstein-Euler equations that
can be expanded to the first post-Newtonian order.

\begin{proof}[Proof of Theorem \ref{mth}]

Using the harmonic equations
\leqn{harmA}{
\ep\del_t\ufb^{44} = - \del_I\ufb^{4I}  , \AND
\ep\del_t\ufb^{I4} = -\del_I\ufb^{IJ} ,
}
we can write the constraint equations \eqref{con.1} as
\alin{constraint}{
\Delta\ufb^{4k} =
\delta^k_4\rho -\delta^k_I\del_L&\ep\del_t\uf^{LI} + \ep \Bigl[Q^{4k}_{0}(\ep\ufb^{ij},\del_{I}\del_J\ufb^{ij},
\ep\del_I\del_t\ufb^{KL}) \notag \\
&+ Q^{4j}_1(\ep^2\ufb^{ij},\del_I\ufb^{ij},\ep\del_t\uf^{IJ})+
Q^{4j}_2(\ep^2\ufb^{ij},\wv,\ep \wv)\alpha^2
\Bigr],
}
where $Q^{4j}_{0}(y_1,y_2,y_3)$ is bilinear in $y_1$ and $(y_2,y_3)$,
$Q_{1}^{4j}(y_1,y_2,y_3)$ is quadratic in $y_2,y_3$, and
the maps $Q_\nu^{4k}$ $(\nu=0,1,2)$ are analytic in all their variables
for $\ep^2\ufb^{ij}\in \Vc$. We can also write the
 $KL$-components
of the reduced Einstein equations \eqref{fo11}
as
\lalign{IJred}{
\del_t^2\ufb^{KL} = \frac{1}{\ep^2(1-\ep^2\ufb^{44})}& \biggl[
\Delta\ufb^{KL}
+2\ep^3\ufb^{I4}\del_I\del_t\ufb^{KL} + \ep^2\ufb^{IJ}\del_{IJ}\ufb^{KL}
+ \ep^2Q_{0}^{KL}(\ep^2\ufb^{ij},
\del_M\ufb^{ij},\ep\del_t\ufb^{IJ})  \notag \\
&   - \ep^2\Bigl(\rho w^Kw^L
+ p\delta^{KL}\Bigr) + \ep^3 Q_1^{KL}(\ep\ufb^{ij},\ep^2\ufb^{ij},
\wv,\ep\wv)\biggr] \, , \label{IJred.1}
}
where
$Q_0^{KL}(y_1,y_2,y_3)$ is quadratic in $(y_2,y_3)$,
\eqn{IJredA}{
Q_1^{KL} = Q_2^{KL}(\ep\ufb^{ij},\ep^2\ufb^{ij},\wv,\ep\wv)\alpha^2 +
Q_3^{KL}(\ep\ufb^{ij},\ep^2\ufb^{ij},\wv,\ep\wv)w^Iw^J\, ,
}
and all of the maps $Q_{\nu}^{KL}$ $(\nu = 0,1,2,3)$ are
analytic in their arguments for $\ep^2\ufb \in \Vc$.

We now take
\eqn{idata}{
\left\{\del_t\ufb^{IJ}(0)=\ep^2\zf_4^{IJ},\alpha(0)=\uset{0}{\alpha},
w^I(0)= \uset{0}{w}^{I}, \ff^{IJ} \right\}
} as the prescribed initial data, and solve the non-linear elliptic system
\lalign{init}{
\Delta\ufb^{4k} &= \Lambda^{4k} :=
\delta^k_4\rho -\delta^k_I\del_L\ep\del_t\uf^{LI} + \ep \Bigl[Q^{4k}_{0}(\ep\ufb^{ij},\del_{I}\del_J\ufb^{ij},
\ep\del_I\del_t\ufb^{KL})
\notag \\
& + Q^{4j}_1(\ep^2\ufb^{ij},\del_I\ufb^{ij},\ep\del_t\uf^{IJ})+
Q^{4j}_2(\ep^2\ufb^{ij},\wv,\ep \wv)\alpha^2
\Bigr], \label{init.1} \\
\Delta\ufb^{KL} &= \Lambda^{KL} :=
-2\ep^3\ufb^{I4}\del_I\del_t\ufb^{KL} - \ep^2\ufb^{IJ}\del_{IJ}\ufb^{KL}
- \ep^2Q_{0}^{KL}(\ep^2\ufb^{ij},
\del_M\ufb^{ij},\ep\del_t\ufb^{IJ}) \notag \\
 &  + \ep^2\Bigl(\rho w^Kw^L
+ p\delta^{KL}\Bigr) - \ep^3 Q_1^{KL}(\ep\ufb^{ij},\ep^2\ufb^{ij},
\wv,\ep\wv) + \ep^4(1-\ep^2\ufb^{44})\ff^{KL},\label{init.2}
}
to determine the initial data $\{\ufb^{ij}|_{t=0},\del_t\ufb^{ij}|_{t=0}\}$ on $S_0$$=$
$\{(x^I,0)\,|\, (x^I)\in\Rbb^3\}$.
Note that $w^4$ is determined by  the fluid velocity normalization
\eqref{con.3},
which can be written as
\leqn{fnorm}{
w^4 = \frac{1}{\ep}f(\ep w^I,\ep^2 \ufb^{ij})\,  ,
}
where  $f(y_1,y_2)$ is analytic in a neighborhood of
$(0,0)$ and $f(\yv) = \text{O}(|\yv|^2)$ as $\yv\rightarrow 0$.

Using the weighted multiplication inequality (see \cite{Oli06}, Lemma A.8) and
Lemma \ref{winqE}, it is straightforward to verify that there exists an $\ep_0 >0$ such that
$\Lambda^{ij}$ (see \eqref{init.1}-\eqref{init.2})
defines an analytic map
\eqn{Lambdamap}{
\bigl(\ep,\zf_4^{IJ},\uset{0}{\alpha},\uset{0}{w}^{I},\ff^{IJ},\ufb^{ij}\bigr)
\in (-\ep_0,\ep_0)\times H^{k}_{\delta-1}\times H^{k-1}_{\delta-1}
\times H^{k}_{\delta-1} \times H^{k-1}_{\delta-2}\times H^k_{\delta}
\longrightarrow \Lambda^{ij} \in H^{k-2}_{\delta-2} \, ,
}
where
\leqn{initB}{
\Lambda^{4i} = \delta^k_4\rho+ \text{0}(\ep) \AND
\Lambda^{KL} = \text{0}(\ep^2)\quad \text{as $\ep\searrow 0$.}
}
Writing \eqref{init.1}-\eqref{init.2} as
\eqn{initA}{
\ufb^{ij} = \Delta^{-1}\Lambda\bigl(\ep,\zf_4^{IJ},\uset{0}{\alpha},\uset{0}{w}^{I},\ff^{IJ},\ufb^{ij}\bigr),
}
it follows from \eqref{initB} and the invertibility of the Laplacian $\Delta\, :\, H^k_\delta \rightarrow H^{k-2}_{\delta-2}$ that we can use the analytic version of the implicit function theorem \cite{Deim}
to conclude that there exists an open neighborhood $U$ of any point in $H^{k}_{\delta-1}\times H^{k}_{\delta-1}
\times H^{k}_{\delta-1} \times H^{k-2}_{\delta-2}$,
and analytic maps
\eqn{solmap}{
\bigl(\ep,\zf_4^{IJ},\uset{0}{\alpha},\uset{0}{w}^{I},\ff^{IJ}\bigr)
\in (-\ep_0,\ep_0)\times U
\longrightarrow \ufb^{ij} \in H^{k}_{\delta}
}
that solve equations \eqref{init.1}-\eqref{init.2}. Moreover, it follows from \eqref{initB} that
\leqn{uKLep0}{
\norm{\ufb^{KL}_\ep(0)}_{H^k_{\delta-1}} \lesssim \ep^2\,  \quad \forall \; \ep \in [0,\ep_0],
}
and hence
\leqn{uKLep2}{
\norm{\del^2_t\ufb^{KL}_\ep(0)}_{H^k_{\delta-2}} \lesssim \ep^2 \quad \forall \; \ep \in [0,\ep_0]\, .
}
Also, we note that by construction
\leqn{uKLep1}{
\norm{\del_t \ufb^{KL}_\ep(0)}_{H^{k-1}_{\delta-1}} \lesssim \ep^2 \quad  \forall \ep \in [0,\ep_0].
}
Differentiating the harmonic conditions \eqref{harmA} with respect to $t$, and
using \eqref{uKLep0}-\eqref{uKLep1}, yields
\lalign{u4jep}{
\norm{\del_t^p\ufb^{44}_\ep(0)}_{H^{k-p}_{\delta-p}} &\lesssim 1 \quad
p=0,\ldots,4\, , \label{u4jep.1}
\intertext{and}
\norm{\del_t^p\ufb^{4J}_\ep(0)}_{H^{k-p}_{\delta-p}} & \lesssim \ep
\quad p= 0,\ldots,3 \label{u4jep.2}
}
for all $\ep \in [0,\ep_0]$.

Using \eqref{harmA}, the Euler
equations \eqref{eul13} can be
written as
\lalign{Euler}{
\del_t \wv = \bigl[a^4(\ep^2\ufb^{ij},\ep\wv)\bigr]^{-1}&\bigl(a^I(w,\ep^2\ufb^{ij},\ep\wv)
\del_I\wv + b_0(\del_I\ufb^{ij},\ep\del_t \ufb^{I4})+ \notag \\
& b_1\bigl(\wv,\ep^2\ufb^{ij},\ep\wv,
\del_I\ufb^{ij},\ep\del_t\ufb^{IJ},\ep\del_I\ufb^{ij},
\ep^2\del_t\ufb^{IJ}\bigr) \bigr) \, , \label{Euler.1}
}
where the maps $a^4$, $a^I$, $b_0$, $b_1$ are analytic in all their arguments
for $\ep^2\ufb\in \Vc$, and $a^4(0,0)=\id$, $a^I(0,0,0,0)=0$,
$b_0(y_1,y_2)$ is linear, and $b^4(y_1,y_2,y_3,y_4,y_5,y_6,y_7)$
is linear in $(y_4,y_5,y_6,y_7)$ and satisfies
$b^4(0,0,0,y_4,y_5,y_6,y_7)=0$.  Then differentiating \eqref{harmA}, \eqref{IJred.1}, and \eqref{Euler.1}
with respect to t while using \eqref{uKLep0}-\eqref{u4jep.2} shows that
\lalign{tdiff}{
\norm{\del_t^p\ufb^{KL}_\ep(0)}_{H^{k-p}_{\delta-2}} &\lesssim 1 \quad p=3,4, \label{tdiff.1} \\
\norm{\del_t^p\alpha_\ep(0)}_{H^{k-p}_{\delta-1}} &\lesssim 1 \quad  p=0,\ldots,3,  \label{tdiff.2} \\
\norm{\del_t^p w^i_\ep(0)}_{H^{k-p}_{\delta-1}} &\lesssim 1 \quad p=0,\ldots,3, \label{tdiff.3}
\intertext{and}
\norm{\ep \del_t^4\ufb^{4J}}_{H^{k-4}_{\delta-3}} &\lesssim 1  \label{tdiff.4}
}
for all $\ep \in [0,\ep_0]$. We then find from the definition of $W_\ep$, the estimates
\eqref{uKLep0}-\eqref{u4jep.2}, and \eqref{tdiff.1}-\eqref{tdiff.4}, that
\leqn{Wtdiff}{
\norm{\del_t^3 W_\ep(0)}_{H^{k-p}_{\delta-1}} \lesssim 1 \quad \text{for $p=0,1,2,3$ and $0\leq \ep \leq \ep_0$.}
}
Next, we observe that
\leqn{L2est}{
\norm{\ufb^{ij}_\ep(t)}_{L^2_\delta} = \norm{
\ufb^{ij}_\ep(0)+ \ep^{-1}\delta\uf^{ij}_\ep(t)}_{L^2_\delta}
\lesssim \norm{\ufb^{ij}_\ep(0)}_{L^2_\delta} + \frac{1}{\ep} \norm{\delta
\ufb^{ij}_\ep(t)}_{L^2_{\delta-1,\ep}}
}
by \eqref{imbedC} and \eqref{imbedD}, while
for any $0\leq \ell \leq k$,
\leqn{Vest}{
\norm{V_\ep(t)} \lesssim
\norm{V_\ep(t)}_{H^\ell_{\delta-1,\ep}}=
\norm{W_\ep(t) + d\Phi(W_\ep(t))}_{H^\ell_{\delta-1,\ep}}
\lesssim \norm{W_\ep(t)}_{H^\ell_{\delta-1,\ep}}}
by \eqref{imbedB}, \eqref{Wdef}, and \eqref{dPhi1}.
The proof of Theorem \ref{mth}, now follows directly
from Theorem \ref{hoeA}, and the estimates
\eqref{Wtdiff}-\eqref{Vest}.
\end{proof}

\sect{disc}{Discussion}

In this article, we have established the existence of a large class
of dynamical solutions to the Einstein-Euler equations that have
a first post-Newtonian expansion. Although this is an improvement
over existing rigorous results \cite{Oli06,Ren94}, which only
cover the Newtonian limit situation (i.e.
the ``zeroth'' post-Newtonian expansion), the results of this paper
are almost certainly not optimal. In general, one expects that
with a suitable gauge choice, it should be possible to generate
post-Newtonian expansions to at least the
$2.5$ post-Newtonian order after which there are indications that
the post-Newtonian expansions will break down. For a lucid
discussion of this phenomenon see \cite{Ren92a}.

As remarked in \cite{Ren92a}, the choice of harmonic
gauge may be the reason for not being able
to reach the $2.5$ post-Newtonian order.  At the formal level, there
exist other gauges that perform better than the
harmonic gauge for the post-Newtonian expansions. However, it
remains to be seen if these other gauges are compatible with
the singular hyperbolic energy estimates that are guaranteed
to arise in the dynamical setting. We are presently investigating
this problem.

From the proof of Theorem \ref{mth} and the
paper \cite{Oli06}, it is clear that conditions of the form
\leqn{disc1}{
\text{$\norm{\del_t^p W_\ep(0)}_{H^{k-p}_{\delta-1}}\lesssim 1$
as $\ep \searrow 0$}
}
on the initial data play a
crucial role in
generating the post-Newtonian expansions. This leads
to the question of what happens when one considers
initial data that does not satisfy \eqref{disc1} for
any $p\in \Zbb \geq 0$. In \cite{Oli08}, we address
this question for the situation where
\eqn{disc2}{
\limsup_{\ep\searrow 0}\norm{\del_t W_\ep(0)}_{H^{k-1}_{\delta-1}}=\infty.
}
There we find that a Newtonian description is still appropriate for
the motion of the matter, but the gravitational field no longer vanishes in
the limit $\ep\searrow 0$. Instead, there exists high frequency gravitational
radiation that is not small at the $\ep^0$ order, and this
will necessarily affect the higher order expansions.

\bigskip

\noindent {\bf Acknowledgements}

\smallskip

\noindent This work began while I was a junior scientist at the Albert-Einstein-Institute (AEI).
I thank the AEI and the director Gerhard Huisken of the Geometric Analysis and Gravitation Group for supporting this research.

\appendix
\sect{winq}{Weighted calculus inequalities}

In this section, we prove additional weighted calculus
inequalities that are similar in spirit to those in Appendix
A of \cite{Oli06}. We first recall from \cite{Oli06} the
definition of the weighted Sobolev spaces.
Let $V$ be a finite dimensional vector space with inner product
$\ipe{\cdot}{\cdot}$ and corresponding norm $\enorm{\cdot}$. For
$u\in L^p_{\text{loc}}(\Rbb^n,V)$, $1\leq p \leq \infty$, $\delta
\in \Rbb$, and $\ep \in \Rbb_{\geq 0}$,  the \emph{weighted
$L^{p}$ norm} of $u$ is defined by \leqn{wLpdef}{
\norm{u}_{L^{p}_{\delta, \ep}} := \left\{\begin{array}{ll}
\norm{\sigma_\ep^{-\delta-n/p}\,u}_{L^p} &
\text{if $ 1\leq p < \infty$}\\
\\
\norm{\sigma_\ep^{-\delta}\,u}_{L^{\infty}} & \text{if $p=\infty$}
\end{array} \right. }where $\displaystyle{\sigma_{\ep}(x) := \sqrt{1+\frac{1}{4}|\ep x|^2}}$. The
\emph{weighted Sobolev norms} are then defined by \leqn{wSobdef}{
\norm{u}_{W^{k,p}_{\delta,\ep}} := \left\{
\begin{array}{ll} \displaystyle{\Bigl(\sum_{|\Ic|\leq k}
\norm{\Der^{\Ic}u}^{p}_{L^{p}_{\delta-|\Ic|,\ep} } \Bigr)^{1/p}} &
\text{if $1\leq p < \infty$} \\
\\ \displaystyle{\sum_{|\Ic|\leq k}
\norm{\Der^{\Ic}u}_{L^{\infty}_{\delta-|\Ic|,\ep} }} & \text{if
$p=\infty$}
\end{array}\right. } where $k\in \Nbb_0$, $\Ic = (\Ic_{1},\ldots, \Ic_{n}) \in
\Nbb_{0}^{n}$ is a multi-index and $\Der^{\Ic} =
\partial_{1}^{\Ic_{1}}\ldots\partial_{n}^{\Ic_{n}}$. Here
\eqn{pardef}{\partial_i = \frac{\partial\;}{\partial x^i}} where
$(x^1,\ldots,x^n)$ are the standard Cartesian coordinates on
$\Rbb^n$.
The weighted Sobolev spaces are then defined as \eqn{wsobdef}{
W^{k,p}_{\delta,\ep} = \{\, u \in W^{k,p}_{\text{loc}}(\Rbb^n,V)
\, | \, \norm{u}_{W^{k,p}_{\delta,\ep}} < \infty \,  \}\, .  }
We note that $W^{k,p}_{\delta,0}$ are the standard Sobolev spaces,
and for $\ep > 0$ the $W^{k,p}_{\delta,\ep}$ are equivalent to
the radially weighted Sobolev spaces \cite{Bart86,CBC}. For $p=2$,
we use the alternate notation $H^{k}_{\delta,\ep} :=
W^{k,2}_{\delta,\ep}$. The spaces $L^{2}_{\delta,\ep}$ and
$H^{k}_{\delta,\ep}$ are Hilbert spaces with inner products
\leqn{L2ip}{ \ip{u}{v}_{L^{2}_{\delta,\ep}} := \int_{\Rbb^{n}}
\ipe{u}{v} \sigma_\ep^{-2\delta-n}d^{n}x\, ,} and \leqn{Hkip}{
\ip{u}{v}_{H^{k}_{\delta,\ep}} := \sum_{|\Ic|\leq k}
\ip{\Der^{\Ic}u}{\Der^{\Ic}v}_{L^{2}_{\delta-|\Ic|,\ep} }, }
respectively. When $\ep =1 $, we will also use the notation
$W^{k,p}_{\delta} = W^{k,p}_{\delta,1}$ and $H^k_\delta =
H^k_{\delta,1}$.

\begin{lem} \label{winqCa} \mnote{[winqCa]}
Suppose $\ep_0>0$, $\delta_1 \geq \max\{\delta_2+\delta_3,
\delta_4+\delta_5\}$, then
\eqn{winCa.1}{
\norm{uv}_{H^k_{\delta_1,\ep}} \lesssim
\norm{u}_{L^\infty_{\delta_2,\ep}}\norm{v}_{H^k_{\delta_3,\ep}}
+ \bigl(\norm{Du}_{H^{k-1}_{\delta_4-1,\ep}}+\ep\norm{u}_{L^2_{\delta_4,\ep}}
\bigr)\norm{v}_{L^\infty_{\delta_5,\ep}}
}
for all $\ep \in [0,\ep_0]$, $u\in L^\infty_{\delta_2,\ep}
\cap H^{k}_{\delta_4,\ep}$, and $v\in L^\infty_{\delta_5,\ep}\cap
H^k_{\delta_3,\ep}$.
\end{lem}
\begin{proof}
This follows directly from the inequality
\eqn{winCa}{
\norm{uv}_{H^k}\lesssim \norm{u}_{L^\infty}\norm{v}_{H^k}
+ \norm{Du}_{H^{k-1}}\norm{v}_{L^\infty}
}
and Lemma A.4 of \cite{Oli06}.
\end{proof}

\begin{lem} \label{winqCb}\mnote{[winqCb]}
Suppose $\ep_0 > 0$, $\delta \leq 0$, $-n/2 \leq \lambda \leq
-n/2+1$, $\lambda \geq \delta$, $k>n/2$, and
$f\in C^k_b(\Rbb^L\times \Rbb^N,\Mbb^{M\times M})$ with
$f(0,0)=0$.
Then there exists a polynomial $p(y_1,y_2,y_3)$ such that
\eqn{winqCb.1}{
\norm{f(u,w)v}_{H^k_{\delta,\ep}} \lesssim \norm{f}_{C^{k}_b}
p\bigl(\norm{u}_{H^k_\lambda},\norm{w}_{H^k_{\delta,\ep}},
\norm{v}_{H^k_{\delta,\ep}} \bigr)\norm{v}_{H^k_{\delta,\ep}}
}
for all $\ep \in [0,\ep_0]$, $u\in H^k_{\lambda}$ and $w,v \in H^k_{\delta,\ep}$.
\end{lem}
\begin{proof}
Since $\delta \leq \lambda$, it follows from
Lemma \ref{winqCa}
that
\eqn{winqCb.2}{
\norm{f(u,w)v}_{H^k_{\delta,\ep}} \lesssim
\norm{f(u,w)}_{L^\infty}\norm{v}_{H^k_{\delta,\ep}}
+ \bigl(\norm{D(f(u,w))}_{H^{k-1}_{\lambda-1,\ep}}+
\ep\norm{f(u,w)}_{H^k_{\lambda,\ep}}\bigr)\norm{v}_{L^\infty_{\delta,\ep}}\, .
}
Using Lemma A.9 of \cite{Oli06}, we can write the above
inequality as
\lalign{winqCb.3}{
\norm{f(u,w)v}_{H^k_{\delta,\ep}}  \lesssim
\norm{f(u,w)}_{L^\infty}\norm{v}_{H^k_{\delta,\ep}}
+ \norm{f}_{C^{k}_{b}}\bigl[1+&\bigl(\norm{u}_{L^\infty}+
\norm{w}_{L^\infty}\bigr)^{k-1}\bigr]\bigl(\norm{Du}_{H^{k-1}_{\lambda-1,\ep}}
\notag \\
+ \ep\norm{u}_{H^{k}_{\lambda,\ep}} +
\norm{w}_{H^{k-1}_{\lambda,\ep}}\bigr) \norm{v}_{L^\infty_{\delta,\ep}} \, .
\label{winqCb.3.1}
}

But $k>n/2$ and $\lambda \leq \delta \leq 0$ implies that
\lgath{winqCb.4}{
\norm{u}_{L^\infty}\lesssim \norm{u}_{H^k_\lambda}\, ,\quad
\norm{w}_{L^\infty} \lesssim \norm{w}_{H^k_{\delta,\ep}}\, ,\quad
\norm{v}_{L^\infty} \lesssim \norm{v}_{L^\infty_{\delta,\ep}} \lesssim \norm{v}_{H^k_{\delta,\ep}} \, , \label{winqCb.4.1}
\intertext{and}
\norm{w}_{H^{k}_{\lambda,\ep}} \lesssim \norm{w}_{H^k_{\delta,\ep}}
\label{winqCb.4.2}
}
by equation A.24 and Lemma A.7 of \cite{Oli06}, while
\leqn{winqCb.5}{
\norm{Du}_{H^{k-1}_{\lambda-1,\ep}}+\ep\norm{u}_{L^2_{\lambda,\ep}}
\lesssim \norm{u}_{H^k_\lambda}
}
follows from Lemma A.11 of \cite{Oli06} since $-n/2\leq \lambda
\leq -n/2+1$. The proof now follows directly from the
inequalities \eqref{winqCb.3.1}-\eqref{winqCb.5}.
\end{proof}

\begin{lem} \label{winqCc} \mnote{[winqCc]}
Suppose $\ep_0 > 0$, $\delta \leq 0$, $-n/2 \leq \lambda \leq
-n/2+1$, $\lambda \geq \delta$, $k>n/2+1$, and
$f\in C^k_b(\Rbb^L\times \Rbb^N,\Mbb^{M\times M})$ with
$f(0,0)=0$.
Then there exists a polynomial $p(y_1,y_2)$ such that
\eqn{winCc.1}{
\norm{[D^\alpha,f(u,w)]v}_{L^2_{\delta-|\alpha|}} \lesssim
\norm{f}_{C^{k}_b}p(\norm{u}_{H^k_\lambda},\norm{w}_{H^k_{\delta,\ep}})
\bigl(\norm{u}_{H^k_\lambda}+\norm{w}_{H^k_{\delta,\ep}}\bigr)\norm{v}_{H^{k-1}_{\delta-1,\ep}}
}
for all $\ep \in [0,\ep_0]$ , $1\leq |\alpha|\leq k$, $u\in H^k_{\lambda}$,
$w\in H^k_{\delta,\ep}$, and $v\in H^{k-1}_{\delta-1,\ep}$.
\end{lem}
\begin{proof}
The proof follows directly  from Lemma A.9 of \cite{Oli06}
and the inequalities \eqref{winqCb.3.1}-\eqref{winqCb.5}.
\end{proof}

\begin{lem} \label{winqCd} \mnote{[winqCd]}
Suppose $\ep_0 > 0$, $\delta \leq 0$, $-n/2 \leq \lambda \leq
-n/2+1$,  $\lambda \geq \delta$,  and $k>n/2$. Then there exists a constant $C>0$ such that
\alin{winqCd1}{
\norm{u_1u_2}_{H^k_\lambda}  & \leq C\norm{u_1}_{H^k_\lambda}\norm{u_2}_{H^k_\lambda}\, , \\
\norm{u_1v_1}_{H^k_{\delta,\ep}} & \leq C\norm{u_1}_{H^k_\lambda}\norm{v_1}_{H^k_{\delta,\ep}} \, ,
\intertext{and}
\norm{v_1v_2}_{H^k_{\delta,\ep}} & \leq C \norm{v_1}_{H^k_{\delta,\ep}} \norm{v_2}_{H^k_{\delta,\ep}}
}
for all $u_1,u_2\in  H^k_{\lambda}$,  $v_1,v_2\in H^k_{\delta,\ep}$, and $\ep \in [0,\ep_0]$.
\end{lem}
\begin{proof}
The proof follows immediately from Lemma \ref{winqCa} and the inequalities
\eqref{winqCb.4.1}-\eqref{winqCb.5}.
\end{proof}

We now recall the definition of analytic maps between Banach spaces.
\begin{Def} \label{winqDa} \mnote{[winqDa}
{\rm Suppose $Y$ and $Z$ are Banach spaces, $U \subset Y$ is an open set, and
$\Lc_j(Y,Z)$ is the space of continuous, $j$-multilinear maps from $Y$ to $Z$
with norm
\eqn{winqDa1}{
\norm{F}_{\Lc_j(Y,Z)} = \sup\,\bigl\{ \norm{F(u_1,u_2,\ldots,u_j)}_{Z} \,\bigl| \,
u_j\in U \AND
\sup\{\norm{u_1}_{Y},\norm{u_2}_{Y},\ldots,\norm{u_3}_{Y}\}\leq 1 \bigr\} .
}
Then
a map $f : U \longrightarrow Z$ is \emph{analytic in U}, if for each $u_0\in U$ there
exists a $\rho > 0$, and a sequence of maps multilinear maps
$f_j \in \Lc_j(Y,Z)$ such that
\eqn{winDa2}{
\sum_{j=0}^\infty \norm{f_j}_{\Lc_j(Y,Z)} \rho^j < \infty,
}
and
\leqn{winDa3}{
f(u) = \sum_{j=0}^\infty f_j(u-u_0,\ldots,u-u_0)
}
for all $u\in U$ satisfying $\norm{u-u_0}_Y < \rho$. The set of all analytic functions
in U will be denoted $C^\omega(U,Z)$.
}
\end{Def}
In addition to analytic maps, we will need analytic maps that are uniformly analytic
on the $H^k_{\delta,\ep}$ spaces as $\ep$ varies.
\begin{Def} { \rm \label{winqD} \mnote{[winqD]}
Suppose $R>0$,  $Y,Z$ are Banach spaces, and $V \subset Y$ is open. Then a sequence a  maps
$f_\ep : B_R(H^{k_1}_{\ep}) \times V
\rightarrow H^{k_2}_{\delta_2,\ep} \times Z$
will be called \emph{uniformly analytic for $\ep \in [0,\ep_0]$}, if
\begin{itemize}
\item[(i)] $f_\ep \in C^\omega( B_{R}(H^{k_1}_{\delta_1,\ep} \times V;H^{k_2}_{\delta_2,\ep} \times Z)$
for $0\leq \ep \leq \ep_0$, and
\item[(ii)] for each $v_0\in V$ there exists constants  $\rho, c_j > 0$, and a sequence of maps multilinear maps
$f^{\ep}_j \in \Lc_j(H^{k_1}_{\delta_1,\ep}\times Y, H^{k_2}_{\delta_2,\ep} \times Z)$ such that
\eqn{winqD1}{
\norm{f^{\ep}_j}_{\Lc_j(H^{k_1}_{\delta_1,\ep}\times Y, H^{k_2}_{\delta_2,\ep} \times Z)}\leq c_j
\qquad 0\leq \ep \leq \ep_0\, ,
}
\eqn{winqD2}{
\sum_{j=0}^\infty  c_j(\rho+R)^j < \infty,
}
and
\eqn{winqD3}{
f_\ep(u,v) = \sum_{j=0}^\infty f^{\ep}_j(u,v-v_0,\ldots,u,v-v_0)  \qquad 0\leq \ep \leq \ep_0
}
for all $(u,v)\in H^{k_1}_{\delta_1,\ep} \times V$ satisfying $\norm{u}_{H^{k_1}_{\delta_1,\ep}} < R$, and $\norm{v-v_0}_Y < \rho$.
\end{itemize}
}
\end{Def}
The next lemma shows how to construct a particular class of uniformly analytic functions.
\begin{lem} \label{winqE} \mnote{[winqE]}
Suppose $\ep_0 > 0$, $\delta \leq 0$, $-n/2 \leq \lambda \leq
-n/2+1$, $k>n/2$, $F \in C^{\omega}(B_{R_1}(\Rbb)\times B_{R_2}(\Rbb),\Rbb)$,
$F(\cdot,0)=0$, and $C$ is the $\ep$ independent
constant from Lemma \ref{winqCd}. Then for $0\leq \ep \leq \ep_0$,
\eqn{winqG.1}{
F(u,v) = \sum_{p=0}^\infty\sum_{q=1}^\infty
\frac{1}{q!p!}\bigl(\del_1^p\del_2^q F\bigr)(0,0) u^p v^{q}
}
defines a function of class $C^{\omega}(B_{\bar{R}_1}(H^k_\lambda)\times B_{\bar{R}_2}(H^k_{\delta,\ep}),H^k_{\delta,\ep})$
where $\bar{R}_1 = R_1/C$ and $\bar{R}_2 = R_2/C$.
\end{lem}
\begin{proof}
Using Lemma \ref{winqCd}, the proof follows from a slight modification of the proof of Proposition 3.6 from
\cite{Heil}.
\end{proof}
We note that the above Lemma can be easily generalized to maps $f \in C^{\omega}(B_{R}(\Rbb^N)
\times B_{R}(\Rbb^M),\Mbb_{M\times M} )$.

\sect{hyp}{Symmetric hyperbolic equations}

The hyperbolic equations that we will consider are of the form
\lalign{hb}{ &b^0(\ep u_\ep,\ep w_\ep,\ep v_\ep)\partial_{t}v_\ep =
\frac{1}{\ep} c^j\del_j v_\ep + b^j(\ep,u_\ep,w_\ep,v_\ep)\partial_j v_\ep +
\gamma F(\ep,u_\ep,w_\ep, v_\ep)
, \label{hb.1}\\
&v_\ep|_{t=0}  = \vo_\ep , \label{hb.2}}
where
\begin{itemize}
\item[(i)] the maps $u_\ep=u_\ep(x)$ and $w_\ep=w_\ep(t,x)$ are $\Rbb^L$ and $\Rbb^N$
valued, respectively, while
the map
$v_\ep=v_\ep(t,x)$ is $\Rbb^{M}$-valued,
\item[(ii)] $F$ is a (possibly non-local) map satisfying
\leqn{Fcond1}{
\norm{F(\ep,u,w_1,v_1)-F(\ep,u,w_2,v_2)}_{H^{k}_{\delta}}
\lesssim_{\rho,\ep_0,k,\ell} \norm{w_1-w_2}_{H^k_{\delta}}+ \norm{v_1-v_2}_{H^{k}_{\delta}}
}
for all $\ep \in [0,\ep_0]$, $u \in B_\rho(H^{\ell}_{\lambda})$, $w_1,w_2,v_1,v_2\in B_\rho(H^{k}_{\delta,\ep})$, and
\leqn{Fcond2}{
\norm{F(\ep,u,w,v)}_{H^k_{\delta,\ep}}\lesssim
p(\norm{u}_{H^\ell_\lambda},\norm{w}_{H^k_{\delta,\ep}},\norm{v}_{H^k_{\delta,\ep}})\bigl(
\norm{w}_{H^k_{\delta,\ep}} + \norm{v}_{H^k_{\delta,\ep}}\bigr)
}
for all $\ep \in [0,\ep]$, $u\in H^\ell_\lambda$, and $w,v\in H^k_{\delta,\ep}$,
\item[(iii)] $b^0, b^j \in C^{\ell}_b(\Rbb^L\times\Rbb^N\times\Rbb^M, \Mbb_{M\times M})$
 $(j=1,\ldots,n)$,
\item[(iv)] $b^0$ and $b^j$ are symmetric,
\item[ (v)] the $c^j$ are constant symmetric matrices, and
\item[(vi)] there exists a constant $\omega > 0$ such that
\leqn{bb}{ b^0(\xi_1,\xi_2,\xi_3) \geq \omega \id_{\!M\times M}
\quad \text{for all $(\xi_1, \xi_2,\xi_3) \in \Rbb^L\times\Rbb^M\times \Rbb^{M}$.}}
\end{itemize}

Let $[n/2]$ denote the largest integer with $[n/2]\leq n/2$, $k_0 = [n/2]+2$, and
\eqn{Xdef}{
X_{T,s,k,\delta} = \bigcap_{\ell=0}^{s+1} C^\ell([0,T),H^{k-\ell}_{\delta}).
}
\begin{thm}\label{hypA} \mnote{[hypA]}
Suppose $\ep_0 >0$, $T>0$, $s\in \mathbb{N}_0$, $k=k_0+s$, $\delta \leq 0$, $-n/2\leq \lambda \leq -n/2+1$,
\gath{hypA.0}{
\vo_\ep \in H^k_\delta, \quad u_\ep \in H^{k}_{\lambda},
\quad  w_\ep \in X_{T,s,k,\delta} ,\qquad  0<\ep\leq \ep_0 ,
\intertext{and}
\norm{\vo_\ep}_{H^k_{\delta,\ep}}
\leq C_1 ,\quad\norm{w_\ep(t)}_{H^k_{\delta,\ep}}
+\norm{\del_t w_\ep(t)}_{H^{k-1}_{\delta,\ep}}
\leq C_2,\quad  \norm{u_\ep}_{H^k_\lambda}\leq C_3 ,
}
for constants $C_1$, $C_2$, $C_3$, independent of $(t,\ep)\in [0,T)\times (0,\ep_0]$.
Then there exists
a polynomial $p(y_1,y_2,y_3)$
and maps
\eqn{hypA.2}{
v_\ep \in X_{T_\ep,s,k,\delta} \qquad  0<\ep \leq \ep_0 ,
}
such that
for all
$\ep \in (0,\ep_0]$
\begin{itemize}
\item[(i)] $v_\ep(t,x)$  is the unique solution
in  $L^\infty((0,T_*),H^k_\delta)\cap \text{{\rm Lip}}((0,T_{\ep}),H^{k-1}_\delta)$ to
the initial value problem  \eqref{hb.1}-\eqref{hb.2}),
\item[(ii)] if $\limsup_{t\nearrow T_\ep}\norm{v_\ep}_{W^{1,\infty}} < \infty$, then
the solution $v_\ep$ can be extended (uniquely) for time $T^*_\ep \in [T_\ep,T)$,
\item[(iii)] for any constant $K_1>C_1$,
\eqn{hypA.3}{
\norm{v_\ep(t)}_{H^k_{\delta,\ep}} \leq \exp\bigl(K_2(1+\gamma)t\bigr)\left[\norm{v_\ep(0)}_{H^k_{\delta,\ep}}
+ \frac{\gamma C_2}{K_2(1+\gamma)}\right]-\frac{\gamma C_2}{K_2(1+\gamma)} \leq K_1
}
for all $(t,\ep)\in [0,\tilde{T})\times (0,\ep_0]$, where
\eqn{hypA.3b}{
K_2 := p(C_3,C_2,K_1) ,
}
and
\eqn{hypA.3c}{
T_\ep \geq \tilde{T}:= \min\left\{T,\frac{1}{K_2(1+\gamma)}
\ln\left(\frac{K_1 K_2(1+\gamma)+\gamma C_2}{C_1 K_2(1+\gamma)+
\gamma C_2}\right)\right\},
}
\item[(iv)]$\ep \norm{v_\ep(t)}_{H^k_{\delta,\ep}} \lesssim 1$ for all
$(t,\ep)\in [0,\tilde{T})\times (0,\ep_0]$,
\item[(iv)]
and if $\norm{c^j\del_j \vo_\ep}_{H^{k-1}_{\delta,\ep}} \lesssim \ep$,
then $\norm{\del_t v_\ep(t)}_{H^{k-1}_{\delta,\ep}} \lesssim 1$ for all
$(t,\ep)\in [0,\tilde{T})\times (0,\ep_0]$.
\end{itemize}
\end{thm}
\begin{proof}
We will only prove statements (iii)-(v) as (i)-(ii) follow from
a slight modification of arguments in Appendix B of \cite{Oli06}\footnote{The only real
difference is the proof of the  convergence of the Galerkin approximations. For
the non-local problem, one can use the global compact imbedding $H^k_{\delta}\subset H^\ell_{\eta}$
$(k>\ell, \delta < \eta)$ to obtain convergence instead
of the local compact $H^k(B_R) \subset H(B_R)^\ell$ $(k>\ell)$ imbedding used in \cite{Oli06}.}.

Let $v_\ep^\alpha = D^\alpha v_\ep$, $b^0_\ep = b^0(\ep u_\ep,\ep w_\ep,\ep v_\ep)$,
$b^j_\ep = b^j(\ep,u_\ep,w_\ep,v_\ep)$, and
$F_\ep = F(\ep,u_\ep,w_\ep,v_\ep)$. Then from the evolution equation \eqref{hb.1}, we find that
\leqn{hypA.8}{
\del_t v_\ep = \left(b^0_\ep\right)^{-1}\left[
\frac{1}{\ep}c^j\del_jv_\ep + b^j_\ep \del_j v_\ep + \gamma F_\ep\right]\, .
}
Differentiating this yields
\leqn{hypA.9}{
b^0_\ep \del_t v^\alpha_\ep = \frac{1}{\ep}c^j\del_j v_\ep + b^j_\ep \del_j v_\ep
+ f^\alpha_\ep\, ,
}
where
\leqn{hypA.10}{
f^\alpha_\ep = b^0_\ep\bigl[D^\alpha,\bigl(b_\ep^0\bigr)^{-1}\bigl(
\ep^{-1}c^j + b^j_\ep\bigr)\bigr)\bigr]\del_j v^\alpha_\ep +
\gamma b^0_\ep D^\alpha\bigl(\bigl(b_\ep^0\bigr)^{-1}F_\ep\bigr)\, .
}
Energy estimates (see Lemma 7.1 in \cite{Oli06}) then show
that
\leqn{hypA.11}{
\frac{d\;}{dt} \nnorm{v^\alpha_\ep}^2_{0,\delta,\ep}
\lesssim \bigl(\norm{\text{div}\, b_\ep}_{L^\infty}+ \norm{\vec{c}+\ep\vec{b}}_{L^\infty}\bigr)
\norm{v^\alpha_\ep}_{L^2_{\delta,\ep}} + \norm{f^\alpha_\ep}_{L^2_{\delta,\ep}}
\norm{v^\alpha_\ep}_{L^2_{\delta,\ep}},
}
where $\text{div}\, b_\ep = \del_t b^0_\ep + \del_j b^j_\ep$,
$\vec{c} = (c^1,\ldots,c^n)$, $\vec{b}=(b^1,\ldots,b^n)$, and
\leqn{hypA.12}{
\nnorm{\cdot}_{k,\delta,\ep} := \sum_{|\alpha|\leq k}\ip{D^\alpha(\cdot)}{b^0_\ep
D^\alpha (\cdot)} \, .
}

Since $b^0_\ep = b^0(\ep u_\ep,\ep w_\ep, \ep v_\ep)$, it follows from Lemma \ref{winqCc} that
\lalign{hypA.12a}{
\norm{[D^\alpha,&\bigl(b_\ep^0\bigr)^{-1}\bigl(
\ep^{-1}c^j +  b^j_\ep\bigr)\bigr)\bigr]\del_j v^\alpha_\ep}_{L^2_{\delta,\ep}} \\
&\lesssim p\bigl(\ep\norm{u_\ep}_{H^\ell_\lambda},\ep\norm{w_\ep}_{H^k_{\delta,\ep}} ,\ep\norm{v_\ep}_{H^k_{\delta,\ep}}
\bigr)
\bigl(\norm{u_\ep}_{H^\ell_\lambda}+ \norm{w_\ep}_{H^k_{\delta,\ep}} +\norm{v_\ep}_{H^k_{\delta,\ep}}\bigr)
\norm{v_\ep}_{H^k_{\delta,\ep}}  \label{hypA.12a.1}
}
for some polynomial $p(y_1,y_2,y_3)$. Using this estimate along
with \eqref{Fcond2} and Lemma \ref{winqCb}, we find that
\leqn{hypA.13}{
\norm{f^\alpha_\ep}_{_{L^2_\delta}} \lesssim
p(\norm{u_\ep}_{H^\ell_\lambda},\norm{w_\ep}_{H^k_{\delta,\ep}} ,
\norm{v_\ep}_{H^k_{\delta,\ep}})\bigl(
\norm{w_\ep}_{H^k_{\delta,\ep}}+\norm{v_\ep}_{H^k_{\delta,\ep}}\bigr)
}
for some polynomial $p(y_1,y_2,y_2)$. Combining the two estimates
\eqref{hypA.11} and \eqref{hypA.13}, and summing over
$\alpha$ $(0\leq |\alpha|\leq k)$ yields
\leqn{hypA.13a}{
\frac{d\;}{dt}\nnorm{v_\ep}_{k,\delta,\ep}
\lesssim p(\norm{u}_{H^\ell_{\lambda}},\norm{w_\ep}_{H^k_{\delta,\ep}},\norm{v}_{H^k_{\delta,\ep}})\bigl(
\gamma \norm{w_\ep}_{H^k_{\delta,\ep}}+ (1+\gamma)\norm{v_\ep}_{H^k_{\delta,\ep}}\bigr) \, .
}
But
\eqn{hypA.14}{
\omega\norm{v_\ep}_{H^k_{\delta,\ep}} \leq
\nnorm{v_\ep}_{k,\delta,\ep} \leq \norm{b^0}_{C^0_b}
\norm{v_\ep}_{H^k_{\delta,\ep}},
}
by \eqref{bb}, and so it follows from
\eqref{hypA.13a} and
Gronwall's inequality that for any constant $K_1>C_1$, if we let
$K_2 = p(C_3,C_2,K_1)$, then
\leqn{hypA.15}{
\norm{v_\ep(t)}_{H^k_{\delta,\ep}} \leq \exp\bigl(K_2(1+\gamma)t\bigr)\left[\norm{v_\ep(0)}_{H^k_{\delta,\ep}}
+ \frac{\gamma C_2}{K_2(1+\gamma)}\right]-\frac{\gamma C_2}{K_2(1+\gamma)}
}
for all $t$ such that $\norm{v_\ep(t)}_{H^k_{\delta,\ep}} \leq K_1$. But $\norm{v_\ep}_{W^{1,\infty}}
\lesssim \norm{v_\ep}_{H^k_{\delta,\ep}}$ by Lemma A.7 of \cite{Oli06}, and hence, by the continuation
principle (ii), we see that
\leqn{hypA.16}{
T_\ep \geq \tilde{T}:= \min\left\{T,\frac{1}{K_2(1+\gamma)}
\ln\left(\frac{K_1 K_2(1+\gamma)+\gamma C_2}{C_1 K_2(1+\gamma)+
\gamma C_2}\right)\right\} .
}
Next, differentiating \eqref{hb.1} with respect to
$t$, it is clear that $\del_tv_\ep$ satisfies a linear
equation of the same structure as \eqref{hb.1}, and therefore
the same estimates used to derive \eqref{hypA.15} also show that
there exists
constants $K_2$, $K_3$ such that
\leqn{hypA.17}{
\norm{\del_t v_\ep(t)}_{H^k_{\delta,\ep}} \leq
e^{K_1t}\norm{\del_tv_\ep(0)}_{H^{k-1}_{\delta,\ep}} + K_3 \quad
\forall\; (t,\ep)\in [0,T_*]\times (0,\ep_0] .
}
The proof now follows from the estimates \eqref{hypA.15}-\eqref{hypA.17}.
\end{proof}


\end{document}